\documentclass{article} % For LaTeX2e
\usepackage{iclr2025_conference,times}

\usepackage{amsmath,amsfonts,bm}

\def\eqref#1{equation~\ref{#1}}
\def\1{\bm{1}}

\DeclareMathAlphabet{\mathsfit}{\encodingdefault}{\sfdefault}{m}{sl}
\SetMathAlphabet{\mathsfit}{bold}{\encodingdefault}{\sfdefault}{bx}{n}

\DeclareMathOperator*{\argmin}{arg\,min}

\usepackage{hyperref}
\usepackage{url}

\usepackage[utf8]{inputenc} % allow utf-8 input
\usepackage[T1]{fontenc}    % use 8-bit T1 fonts
\usepackage{hyperref}       % hyperlinks
\usepackage{url}            % simple URL typesetting
\usepackage{booktabs}       % professional-quality tables
\usepackage{amsfonts}       % blackboard math symbols
\usepackage{nicefrac}       % compact symbols for 1/2, etc.
\usepackage{microtype}      % microtypography
\usepackage{xcolor}         % colors
\usepackage{caption} 
\usepackage{graphicx}
\usepackage{booktabs} % for professional tables
\usepackage{amssymb}
\usepackage{amsmath}
\usepackage{multirow}
\usepackage{subcaption}
\usepackage{float}
\usepackage{mathtools}
\usepackage{bbold}
\usepackage{subcaption}

\usepackage[linesnumbered,ruled,vlined]{algorithm2e}
\DontPrintSemicolon

\SetKwComment{Comment}{\color{green!50!black}// }{}

\newcommand{\var}{\texttt}

\SetKwProg{Function}{function}{}{}

\usepackage{pythontex} 
\usepackage{makecell}
\usepackage{wrapfig}

\newcommand{\comment}[1]{}
\DeclareMathOperator{\bin}{bin}
\newcommand{\mathbbm}[1]{\text{\usefont{U}{bbm}{m}{n}#1}} % from mathbbm.sty

\title{Factor Graph Optimization of Error-Correcting Codes for Belief Propagation Decoding}

\author{Yoni Choukroun \\
The Blavatnik School of Computer Science\\
Tel Aviv University\\
\texttt{choukroun.yoni@gmail.com} \\
\And
Lior Wolf \\
The Blavatnik School of Computer Science\\
Tel Aviv University\\
\texttt{wolf@cs.tau.ac.il} \\
}
\iclrfinalcopy % Uncomment for camera-ready version, but NOT for submission.
\begin{document}

\maketitle
%%%%%%%%%%%%%%%%%%%%%%%%%%%%%%%%%%%%%%%%%%%%%
%%%%%%%%%%%%%%%%%%%%%%%%%%%%%%%%%%%%%%%%%%%%%%%%%%%%%%%%%%%%%%%%%%%%%%%%%%%%%%%%%%%%%%%%%%%%%%%%%%%%%%%%%%%%%%%%%%%%%
%%%%%%%%%%%%%%%%%%%%%%%%%%%%%%%%%%%%%%%%%%%%%%%%%%%%%%%%%%%%%%%%%%%%%%%%%%%%%%%%%%%%%%%%%%%%%%%%%%%%%%%%%%%%%%%%%%%%%
%%%%%%%%%%%%%%%%%%%%%%%%%%%%%%%%%%%%%%%%%%%%%%%%%%%%%%%%%%%%%%%%%%%%%%%%%%%%%%%%%%%%%%%%%%%%%%%%%%%%%%%%%%%%%%%%%%%%%
%%%%%%%%%%%%%%%%%%%%%%%%%%%%%%%%%%%%%%%%%%%%%%%%%%%%%%%%%%%%%%%%%%%%%%%%%%%%%%%%%%%%%%%%%%%%%%%%%%%%%%%%%%%%%%%%%%%%%
%%%%%%%%%%%%%%%%%%%%%%%%%%%%%%%%%%%%%%%%%%%%%%%%%%%%%%%%%%%%%%%%%%%%%%%%%%%%%%%%%%%%%%%%%%%%%%%%%%%%%%%%%%%%%%%%%%%%%
%%%%%%%%%%%%%%%%%%%%%%%%%%%%%%%%%%%%%%%%%%%%%%%%%%%%%%%%%%%%%%%%%%%%%%%%%%%%%%%%%%%%%%%%%%%%%%%%%%%%%%%%%%%%%%%%%%%%%

\begin{abstract}
The design of optimal linear block codes capable of being efficiently decoded is of major concern, especially for short block lengths. 
As near capacity-approaching codes, Low-Density Parity-Check (LDPC) codes possess several advantages over
other families of codes, the most notable being its efficient decoding via Belief Propagation.
 While many LDPC code design methods exist, the development of efficient sparse codes that meet the constraints of modern short code lengths and accommodate new channel models remains a challenge.
In this work, we propose for the first time a gradient-based data-driven approach for the design of sparse codes. We develop locally optimal codes with respect to Belief Propagation decoding via the learning of the Factor graph under channel noise simulations. 
This is performed via a novel complete graph tensor representation of the Belief Propagation algorithm, optimized over finite fields via backpropagation and coupled with an efficient line-search method. 
The proposed approach is shown to outperform the decoding performance of existing popular codes by orders of magnitude and demonstrates the power of data-driven approaches for code design.
\end{abstract}

%%%%%%%%%%%%%%%%%%%%%%%%%%%%%%%%%%%%%%%%%%%%%
%%%%%%%%%%%%%%%%%%%%%%%%%%%%%%%%%%%%%%%%%%%%%%%%%%%%%%%%%%%%%%%%%%%%%%%%%%%%%%%%%%%%%%%%%%%%%%%%%%%%%%%%%%%%%%%%%%%%%
%%%%%%%%%%%%%%%%%%%%%%%%%%%%%%%%%%%%%%%%%%%%%%%%%%%%%%%%%%%%%%%%%%%%%%%%%%%%%%%%%%%%%%%%%%%%%%%%%%%%%%%%%%%%%%%%%%%%%
%%%%%%%%%%%%%%%%%%%%%%%%%%%%%%%%%%%%%%%%%%%%%%%%%%%%%%%%%%%%%%%%%%%%%%%%%%%%%%%%%%%%%%%%%%%%%%%%%%%%%%%%%%%%%%%%%%%%%
%%%%%%%%%%%%%%%%%%%%%%%%%%%%%%%%%%%%%%%%%%%%%%%%%%%%%%%%%%%%%%%%%%%%%%%%%%%%%%%%%%%%%%%%%%%%%%%%%%%%%%%%%%%%%%%%%%%%%
%%%%%%%%%%%%%%%%%%%%%%%%%%%%%%%%%%%%%%%%%%%%%%%%%%%%%%%%%%%%%%%%%%%%%%%%%%%%%%%%%%%%%%%%%%%%%%%%%%%%%%%%%%%%%%%%%%%%%
%%%%%%%%%%%%%%%%%%%%%%%%%%%%%%%%%%%%%%%%%%%%%%%%%%%%%%%%%%%%%%%%%%%%%%%%%%%%%%%%%%%%%%%%%%%%%%%%%%%%%%%%%%%%%%%%%%%%%

\section{Introduction}
Reliable digital communication is of major importance in the modern information age and involves the design of codes that can be robustly and efficiently decoded despite noisy transmission channels. 
Over the last half-century, significant research has been dedicated to the study of capacity-approaching Error Correcting Codes (ECC) \citep{shannon1948mathematical}. 
Despite the initial focus on short and medium-length linear block codes \citep{berlekamp1974key}, the development of long channel codes \citep{forney1966concatenated,costello2007channel} has emerged as a viable approach to approaching channel capacity \citep{berrou1993near,mackay1999good,richardson2001design,richardson2001capacity,arikan2008channel,luby2001improved,kudekar2011threshold}. 
  
While the NP-hard maximum likelihood rule defines the target decoding of a given code, developing more practical solutions generally relies on theories grounded upon asymptotic analysis over conventional communication channels. 
However, modern communication systems rely on the design of short and medium-block-length codes \citep{liva2016code} and the latest communication settings provide new types of channels. 
This is mainly due to emergent applications in the modern wireless realm requiring the transmission of short data units, such as remote command links, Internet of Things, and messaging services \citep{de2011reliability,boccardi2014five,paolini2015coded,durisi2016toward, ETSI}.
These challenges call for the formulation of data-driven solutions, capable of adapting to various settings of interest and constraints, generally uncharted by existing theories.

The vast majority of existing machine-learning solutions to the ECC problem concentrate on the design of \emph{neural decoders}. The first neural models focused on the implementation of parameterized versions of the legacy Belief Propagation (BP) decoder \citep{nachmani2016learning, nachmani2017learning,lugosch2017neural,nachmani2019hyper,buchberger2020learned}.
Recently, state-of-the-art learning-based de novo decoders have been introduced, borrowing from well-proven architectures from other domains. 
A Transformer-based decoder that incorporates the code into the architecture has been recently proposed by \citep{choukroun2022error}, outperforming existing methods by sizable margins and at a fraction of their time complexity. 
This architecture has been subsequently integrated into a denoising diffusion models paradigm, further improving results \citep{choukroun2022zdenoising}. 
Subsequently, a universal neural decoder has been proposed in \citep{choukroun2024found}, capable of unified decoding of codes from different families, lengths, and rates. 
Most recently  and related to our work, \citep{choukroun2024colearning} developed an end-to-end learning framework capable of co-learning binary linear block codes along with the neural decoder. 

However, neural decoding methods require increased computational and memory complexity compared to their well-established classical counterparts. Due to these challenges, and the non-trivial acceleration and implementation required, neural decoders were never deployed in real-world systems, as far as we know.

In this work, given the ubiquity and advantages of the Belief Propagation (BP) algorithm \citep{pearl1988probabilistic, richardson2001design} for sparse codes, we consider the {optimization} of codes with respect to BP via the learning of the underlying factor/Tanner graph. 
From a graphical probabilistic model perspective \cite{koller2009probabilistic}, BP being a marginalization algorithm, a gradient-based of a score metric method is given for the \emph{structure learning} of BP's underlying Bayesian network in an end-to-end fashion.
As far as we can ascertain, this is the first time a gradient-based data-driven solution is given for the design of the codes themselves based on a classical decoder. Such a solution induces a very low overhead (if any) for integration into the existing decoding solutions. 

Beyond the conceptual novelty, we make three technical contributions: (i) we formulate the data-driven optimization objective adapted to the setting of interest (e.g., channel noise, code structure),  (ii) we reformulate BP in a tensor fashion to learn the {connectivity} of the factor graph through backpropagation, and (iii) we propose a differentiable and fast optimization approach via a line-search method adapted to the relaxed binary programming setting. 
Applied to a wide variety of codes, our method produces codes that outperform existing codes on various channel noise settings, demonstrating the power and flexibility of the method in adapting to realistic settings of interest.

\section{Related Works}

Neural decoder or data-driven contributions generally focus on short and moderate-length codes for two main reasons.
First, classical decoders reach the capacity of the channel for large codes, and  
second, the emergence of applications driven by the Internet of Things requires effective decoders for short to moderate-length codes. For example, 5G Polar codes have code lengths of 32 to 1024~\citep{liva2016code,ETSI}.

Previous work on neural decoders is generally divided into two main classes: model-free and model-based. Model-free decoders employ general types of neural network architectures \citep{cammerer2017scaling,gruber2017deep,kim2018communication, bennatan2018deep, jiang2019deepturbo, choukroun2022error, choukroun2022zdenoising, choukroun2024found,choukroun2024deep}. Model-based decoders implement parameterized versions of classical Belief Propagation (BP) decoders, where the Tanner graph is unfolded into an NN in which scalar weights are assigned to each variable edge. This results in an improvement in comparison to the baseline BP method for short codes \citep{nachmani2016learning,nachmani2019hyper,raviv2020graph,raviv2023crc,kwak2023boosting}. While model-based decoders benefit from a strong theoretical background, the architecture is overly restrictive, which generally enforces its coupling with high-complexity NN \citep{nachmani2021autoregressive}. Also, the improvement gain generally vanishes for more iterations and longer codewords \citep{sionna} and the integration cost remains very high due to both computational and memory requirements.

% choukroun2024colearning

While neural decoders show improved performance in various communication settings, there has been very limited success in the design of novel neural coding methods, which remain impracticable for deployment \citep{AutoencoderComm,kim2018deepcode,jiang2019turbo}.
Recently, \citep{choukroun2024colearning} provided a new differentiable way of designing binary linear block codes (i.e., parity-check matrices) for a given neural decoder {also showing improved performance with classical decoders}.

Belief-propagation decoding has multiple advantages for LDPC codes \citep{gallager1962low, richardson2001capacity,richardson2001design}.
A large number of LDPC code (parity check matrix) design techniques exist in the literature, depending on the design criterion.
Among them, Gallager \citep{gallager1962low} developed the first regular LDPC codes as the concatenations of permuted sub-matrices. MacKay \citep{mackay1995good} demonstrated the ability of sparse codes to reach near-capacity limits via semi-randomly generated matrices. Irregular LDPC codes have been developed by \citep{richardson2001design,luby2001improved,chung2001design} where the decoding threshold can be optimized via density-evolution.
Progressive Edge Growth \citep{hu2001progressive,hu2005regular} has been proposed to design large girth codes.
Certain classes of LDPC array codes have been presented in \citep{eleftheriou2002low} and LDPC codes with combinatorial design constraints have been developed in \citep{vasic2004combinatorial}.
Finite geometry codes have been developed in \citep{lucas2000iterative, kou2001low} and repeat-accumulate codes have been proposed by \citep{yang2004design,jin2000irregular,narayanaswami2001coded}.
However, the classical methods are not data-driven and are difficult to adapt to the design of codes under constrained settings of interest (e.g., short codes, modern channels, structure constraints, etc.). 
Most related to our work are methods for structure learning \cite{koller2009probabilistic} for Bayesian networks such as the Chow-Liu Algorithm \cite{chow1968approximating} or search-based methods \cite{tian2013branch}.
Related to greedy search-based methods, \cite{elkelesh2019decoder} recently suggested the application of classical genetic algorithms for the discovery of better IRA codes.
\section{Background}
%%%%%%%%%%%%%%%%%%%%%%%%%%%%%%%%%%%%%%%%%%%%%%%%%%%%%%%%%%%%%%%%%%%%%%%%%%%%%%%%%%%%%%%%%%%%%%%%%%%%%%%%%%%%%%%%%%%%%
%%%%%%%%%%%%%%%%%%%%%%%%%%%%%%%%%%%%%%%%%%%%%%%%%%%%%%%%%%%%%%%%%%%%%%%%%%%%%%%%%%%%%%%%%%%%%%%%%%%%%%%%%%%%%%%%%%%%%
%%%%%%%%%%%%%%%%%%%%%%%%%%%%%%%%%%%%%%%%%%%%%%%%%%%%%%%%%%%%%%%%%%%%%%%%%%%%%%%%%%%%%%%%%%%%%%%%%%%%%%%%%%%%%%%%%%%%%
%%%%%%%%%%%%%%%%%%%%%%%%%%%%%%%%%%%%%%%%%%%%%%%%%%%%%%%%%%%%%%%%%%%%%%%%%%%%%%%%%%%%%%%%%%%%%%%%%%%%%%%%%%%%%%%%%%%%%
%%%%%%%%%%%%%%%%%%%%%%%%%%%%%%%%%%%%%%%%%%%%%%%%%%%%%%%%%%%%%%%%%%%%%%%%%%%%%%%%%%%%%%%%%%%%%%%%%%%%%%%%%%%%%%%%%%%%%
%%%%%%%%%%%%%%%%%%%%%%%%%%%%%%%%%%%%%%%%%%%%%%%%%%%%%%%%%%%%%%%%%%%%%%%%%%%%%%%%%%%%%%%%%%%%%%%%%%%%%%%%%%%%%%%%%%%%%
%%%%%%%%%%%%%%%%%%%%%%%%%%%%%%%%%%%%%%%%%%%%%%%%%%%%%%%%%%%%%%%%%%%%%%%%%%%%%%%%%%%%%%%%%%%%%%%%%%%%%%%%%%%%%%%%%%%%%
%%%%%%%%%%%%%%%%%%%%%%%%%%%%%%%%%%%%%%%%%%%%%%%%%%%%%%%%%%%%%%%%%%%%%%%%%%%%%%%%%%%%%%%%%%%%%%%%%%%%%%%%%%%%%%%%%%%%%
%%%%%%%%%%%%%%%%%%%%%%%%%%%%%%%%%%%%%%%%%%%%%%%%%%%%%%%%%%%%%%%%%%%%%%%%%%%%%%%%%%%%%%%%%%%%%%%%%%%%%%%%%%%%%%%%%%%%%
%\subsection{Error Correction Coding}
\textcolor{black}{We assume a standard transmission protocol using a linear block code $C$. The code is defined by a {generator} matrix $G\in \{0,1\}^{k \times n}$ and the parity check matrix $H\in \{0,1\}^{(n - k) \times n}$ is defined such that $GH^{T}=0$ over the order 2 Galois field $GF(2)$.}
{The parity check matrix $H$ entails what is known as a Tanner graph \citep{tanner1981recursive}, which consists of $n$ variable nodes and $(n-k)$ check nodes. The edges of this bipartite graph correspond to the on-bits of the matrix $H$.

The input message $m \in \{0, 1\}^{k}$ is encoded by $G$ to a codeword $c \in C \subset \{0, 1\}^{n}$ satisfying $Hc=H(mG)=0$ and transmitted via a Binary-Input Symmetric-Output channel, e.g., an AWGN channel.
Let $y$ denote the channel output represented as $y=c_{s}+\varepsilon$, where $c_s$ denotes the transmission modulation of $c$ (e.g., Binary Phase Shift Keying (BPSK)), and $\varepsilon$ is random noise independent of the transmitted $c$.
The main goal of the decoder $f_{H}:\mathbb{R}^{n}\rightarrow \mathbb{R}^{n}$ conditioned on the code (i.e., $H$) is to provide a soft approximation $\hat{x}=f_{H}(y)$ of the codeword.

The Belief Propagation algorithm allows the iterative transmission (propagation) of a current codeword estimate (belief) via a Trellis graph determined according to a \emph{factor graph} defined by the code (i.e., the Tanner graph). The factor graph is unrolled into a Trellis graph, initiated with $n$ variable nodes, and composed of two types of interleaved layers defined by the check/factor nodes and variable nodes.
An illustration of the Tanner graph unrolled to the Trellis graph is given in Figure ~\ref{fig:tanner_Trellis}.

As a message-passing algorithm, Belief Propagation operates on the Trellis graph by propagating the messages from {variable nodes} to {check nodes} and from {check nodes} to {variable nodes}, in an alternative and iterative fashion. The input layer generally corresponds to the vector of log-likelihood ratios (LLR) $L \in \mathbb{R}^n$ of the channel output $y$ defined as
$$
L_v = \log\bigg(\frac{\Pr\left(c_v=1 | y_v\right)}{\Pr\left(c_v=0 | y_v\right)}\bigg).
$$
Here, we describe ECC's classical notation of BP with, $v\in \{1,\dots, n\}$ denotes the index corresponding to the $v^{th}$ element of the channel output $y$, for the corresponding bit $c_v$ we wish to recover.  

Let $x^i$ be the vector of messages that a column/layer in the Trellis graph propagates to the next one. 
At the first round of message passing, a variable node type of computation is performed such that

\begin{equation}
x^{2k+1}_e = x^{2k+1}_{(c,v)} = L_v + \sum_{e'\in N(v)\setminus \{(c,v)\}} x^{2k}_{e'}.
\label{eq:base_var}
\end{equation}
Here, each message indexed by the edge $e=(c,v)$ on the Tanner graph and $N(v)=\{(c,v) | H(c,v)=1\}$, i.e, the set of all edges in which $v$ participates. By definition $x^0=0$ such that the messages are directly determined by the vector $L$ for $k=1$. 

For even layers, the check layer performs the following
\begin{equation}
x^{2k}_e = x^{2k}_{(c,v)} = 2\text{arctanh} \left( \prod_{e'\in N(c) \setminus \{(c,v)\}}{\tanh \left ( \frac{x^{2k-1}_{e'}}{2} \right ) }\right)
\label{eq:base_check}
\end{equation}
where $N(c)=\{(c,v) | H(c,v)=1\}$ is the set of edges in the Tanner graph in which row $c$ of the parity check matrix $H$ participates.

The final $v^{th}$ output layer of the BP algorithm, which corresponds to the soft-decision output of the codeword, is given by
\begin{equation}
o_{v} = L_{v} + \sum_{e^{'}\in N(v)} x_{e^{'}}^{2L}
\label{eq:base_check2}
\end{equation}

\begin{figure}[t]
\centering
\includegraphics[width=1\columnwidth]{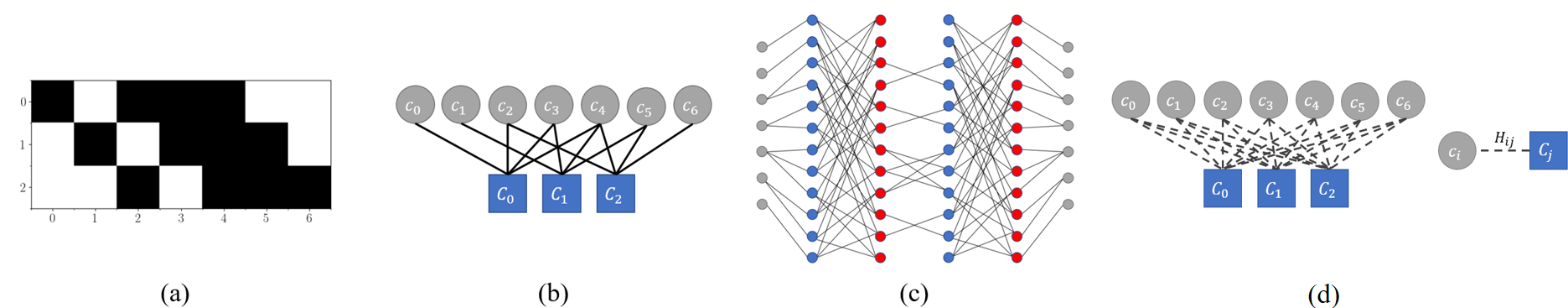}
\caption{For the Hamming(7,4) Code: (a) parity check matrix, the induced (b) Tanner graph, and (c) the corresponding unrolled Trellis graph with two iterations, with odd layers in blue and even layers in red. 
In (d) we present our approach for structure learning via the learned binary weighting of the edges of the \emph{complete} bipartite factor graph unlike the conventional sparse representation.
}
\label{fig:tanner_Trellis}
\end{figure}
\section{Method}
\label{sec:training}
%We present the elements of the proposed design of BP codes as well as the optimization procedure.
%%%%%%%%%%%%%%%%%%%%%%%%%%%%%%%%%%%%%%%%%%%%%%%%%%%%%%%%%%%%%%%%%%%%%%%%%%%%%%%%%%%%%%%%%%%%%%%%%%%%%%%%%%%%%%%%%%%%%
%%%%%%%%%%%%%%%%%%%%%%%%%%%%%%%%%%%%%%%%%%%%%%%%%%%%%%%%%%%%%%%%%%%%%%%%%%%%%%%%%%%%%%%%%%%%%%%%%%%%%%%%%%%%%%%%%%%%%
%%%%%%%%%%%%%%%%%%%%%%%%%%%%%%%%%%%%%%%%%%%%%%%%%%%%%%%%%%%%%%%%%%%%%%%%%%%%%%%%%%%%%%%%%%%%%%%%%%%%%%%%%%%%%%%%%%%%%
%%%%%%%%%%%%%%%%%%%%%%%%%%%%%%%%%%%%%%%%%%%%%%%%%%%%%%%%%%%%%%%%%%%%%%%%%%%%%%%%%%%%%%%%%%%%%%%%%%%%%%%%%%%%%%%%%%%%%
%%%%%%%%%%%%%%%%%%%%%%%%%%%%%%%%%%%%%%%%%%%%%%%%%%%%%%%%%%%%%%%%%%%%%%%%%%%%%%%%%%%%%%%%%%%%%%%%%%%%%%%%%%%%%%%%%%%%%
%%%%%%%%%%%%%%%%%%%%%%%%%%%%%%%%%%%%%%%%%%%%%%%%%%%%%%%%%%%%%%%%%%%%%%%%%%%%%%%%%%%%%%%%%%%%%%%%%%%%%%%%%%%%%%%%%%%%%
%%%%%%%%%%%%%%%%%%%%%%%%%%%%%%%%%%%%%%%%%%%%%%%%%%%%%%%%%%%%%%%%%%%%%%%%%%%%%%%%%%%%%%%%%%%%%%%%%%%%%%%%%%%%%%%%%%%%%
%%%%%%%%%%%%%%%%%%%%%%%%%%%%%%%%%%%%%%%%%%%%%%%%%%%%%%%%%%%%%%%%%%%%%%%%%%%%%%%%%%%%%%%%%%%%%%%%%%%%%%%%%%%%%%%%%%%%%
%%%%%%%%%%%%%%%%%%%%%%%%%%%%%%%%%%%%%%%%%%%%%%%%%%%%%%%%%%%%%%%%%%%%%%%%%%%%%%%%%%%%%%%%%%%%%%%%%%%%%%%%%%%%%%%%%%%%%
%\subsection{Belief Propagation Codes}
The performance of BP is strongly tied to the underlying Tanner graph induced by the code. BP and its variants are generally implemented over a \emph{fixed} sparse graph, such that the only degree of freedom resides in the number of decoding iterations.
While several recent contributions \citep{nachmani2016learning,nachmani2019hyper} aim to enhance the BP algorithm by augmenting the Trellis graph with neural networks, these approaches assume and maintain fixed codes. 
% While several contributions have recently been proposed \citep{nachmani2016learning,nachmani2019hyper} in order to improve the BP algorithm via the augmentation of the Trellis graph with neural networks, the code is assumed and remains fixed.
Here, we propose optimizing the code for the BP algorithm on a decoding setting of interest. Given a trainable binary parity check matrix $H$, we wish to obtain BP-optimized codes by solving the following {\color{black} parameterized} optimization problem

\begin{equation}
\begin{aligned} 
\label{eq:e2e_obj}
H^{*} = \argmin_{H \in \{0,1\}^{(n-k)\times n}} \mathbb{E}_{m\sim \text{Bern}^{k}({\nicefrac{1}{2}}), \varepsilon \sim \mathcal{Z}, T\in \mathbb{N}_{+}} 
% \Big[
&\mathcal{D}\bigg(f_{H,T}\big(\phi(G(H)m)+{\varepsilon}\big), m\bigg) +\mathcal{R}(H)
% \\
% &+\mathcal{R}(P_{\Omega})
% \Big]
\end{aligned}
\end{equation}
Here, $G(H)$ denotes the generator matrix defined by $H$ (i.e., $GH^{T}=0$), $\phi$ denotes the modulation function such that $c_{s}=\phi(c)$, and $\mathcal{Z}$ is the channel noise distribution. 
$f_{H, T}$ denotes the BP decoder built upon $H$ with $T$ iterations (sampled uniformly from a given set), $\mathcal{D}$ denotes the distance metric of interest, and $\mathcal{R}$ denotes the potential {\color{black} hard/soft} regularization of interest,  e.g., sparsity or constraints on the code structure.

Several challenges arise from this optimization problem: 
(i) the optimization is highly non-differentiable and results in an NP-hard binary non-linear integer programming problem, 
(ii) the computation of the codewords $c=Gm$ is both highly non-differentiable (matrix-vector multiplication over $GF(2)$) and computationally expensive (inverse via Gaussian elimination of $H$),  
(iii) the modulation $\phi(\cdot)$ can be non-differentiable, and 
last but most important, (iv) BP assumes a \emph{fixed} code (i.e., the factor graph edges) upon which the decoder is implemented.

%%%%%%%%%%%%%%%%%%%%%%%%%%%%%%%%%%%%%%%%%%%%%%%%%%%%%%%%%%%%%%%%%%%%%%%%%%%%%%%%%%%%%%%%%%%%%%%%%%%%%%%%%%%%%%%%%%%%%
%%%%%%%%%%%%%%%%%%%%%%%%%%%%%%%%%%%%%%%%%%%%%%%%%%%%%%%%%%%%%%%%%%%%%%%%%%%%%%%%%%%%%%%%%%%%%%%%%%%%%%%%%%%%%%%%%%%%%
%%%%%%%%%%%%%%%%%%%%%%%%%%%%%%%%%%%%%%%%%%%%%%%%%%%%%%%%%%%%%%%%%%%%%%%%%%%%%%%%%%%%%%%%%%%%%%%%%%%%%%%%%%%%%%%%%%%%%
%%%%%%%%%%%%%%%%%%%%%%%%%%%%%%%%%%%%%%%%%%%%%%%%%%%%%%%%%%%%%%%%%%%%%%%%%%%%%%%%%%%%%%%%%%%%%%%%%%%%%%%%%%%%%%%%%%%%%
%%%%%%%%%%%%%%%%%%%%%%%%%%%%%%%%%%%%%%%%%%%%%%%%%%%%%%%%%%%%%%%%%%%%%%%%%%%%%%%%%%%%%%%%%%%%%%%%%%%%%%%%%%%%%%%%%%%%%
%%%%%%%%%%%%%%%%%%%%%%%%%%%%%%%%%%%%%%%%%%%%%%%%%%%%%%%%%%%%%%%%%%%%%%%%%%%%%%%%%%%%%%%%%%%%%%%%%%%%%%%%%%%%%%%%%%%%%
%%%%%%%%%%%%%%%%%%%%%%%%%%%%%%%%%%%%%%%%%%%%%%%%%%%%%%%%%%%%%%%%%%%%%%%%%%%%%%%%%%%%%%%%%%%%%%%%%%%%%%%%%%%%%%%%%%%%%
%%%%%%%%%%%%%%%%%%%%%%%%%%%%%%%%%%%%%%%%%%%%%%%%%%%%%%%%%%%%%%%%%%%%%%%%%%%%%%%%%%%%%%%%%%%%%%%%%%%%%%%%%%%%%%%%%%%%%
%%%%%%%%%%%%%%%%%%%%%%%%%%%%%%%%%%%%%%%%%%%%%%%%%%%%%%%%%%%%%%%%%%%%%%%%%%%%%%%%%%%%%%%%%%%%%%%%%%%%%%%%%%%%%%%%%%%%%

\paragraph{Learning the Factor graph via \emph{Tensor} Belief \emph{Back}propagation}
To obtain BP codes, we propose a \emph{structure/Tanner graph learning} approach, where the bipartite graph is assumed as \textbf{complete} with \textbf{learnable} binary-weighted edges. This way, the tensor reformulation of BP weighted by $H$ allows a differentiable optimization of the Tanner graph itself. The two alternating stages of BP can now be represented in a differentiable matrix form rather than its static graph formulation, where the variable layers can be rewritten as 
\begin{equation}
\begin{aligned} 
\label{eq:tensor_eq_var}
Q_{ij}  &=  L_{i} +   \sum_{j'\in C_{i} \setminus j}{R_{j' i}} \equiv L_{i} +   \sum_{j'}{R_{j' i}H_{j'i}} -R_{ji}\,,\\ 
\end{aligned}
\end{equation} }{\color{black} where $R_{ij}$ are the check layers, which are now represented as }
%{\color{black} THIS IS JUST ANOTHER REPRESENTATION OF BP BASED ON THE PCM RATHER THAN THE TANNER GRAPH. IT S WELL KNOWN. EVERYTHING'S DEFINED}
%Similarly, the check layers are now represented as
\begin{equation}
\begin{aligned} 
\label{eq:tensor_eq_var2}
R_{ji}  &=  2\text{arctanh} \left( \prod_{i'\in V_{j} \setminus i}{\tanh \left ( \frac{Q_{i'j}}{2} \right ) }\right) \\
&= 2\text{arctanh} \left( \prod_{i'}\left( {\tanh \left ( \frac{Q_{i'j}H_{ji'}}{2}\right ) +\left( 1-H_{ji'} \right)}\right) / (\tanh \left ( \frac{Q_{ij}}{2} \right)  \right)\,,
\end{aligned}
\end{equation}
% while $R_{ji}$ is further divided by $\tanh \left ( \frac{Q_{ij}}{2} \right) +\epsilon$ to remove the contribution of $Q_{ij}$.
where $C_{i}$ and $V_{j}$ correspond to the non-zero elements in column $i$ and row $j$ of $H$, respectively, while the ones elements in $(1-H)\in \{0,1\}^{(n-k)\times n}$ satisfy the identity element of multiplication.
Potential zero denominators have not been observed but can be handled via regularization or omission.  
As we can observe, BP remains differentiable with respect to $H$ as a composition of differentiable functions.

We provide in Algorithm \ref{algo:bp_tensor} the pseudo-code for the tensor formulation of the BP algorithm, implementing Eq.~\ref{eq:tensor_eq_var2} and~\ref{eq:tensor_eq_var}. 
% {\color{red}ADD LINE BY LINE DESCRIPTION/EXPLNATION OF OF THE CODE, REFERRING TO THE ABOVE EQUATIONS}. 
%{\color{black} It's the exact implementation of the above equations in (pseudo) python.}

\begin{algorithm}[t]
\footnotesize 
  \Function{BP(llr, H, iters, eps=1e-7)}{
   \Comment{llr is the batched LLR matrix, (B, n)}
   \Comment{H is the binary parity-check matrix, (n-k,n)}
   \Comment{iters is the number of BP iterations}
    % \Comment{\color{red}VERY NICE BUT FIRST YOU NEED TO DO A LINE BY LINE EXPLANTION IN THE TEXT ITSELF. SECOND YOU PUT HERE PYTHNE TOGETHER WITH PSUEDOCODE EG IN LINE 2. THIRD, MAYBE PUT SIDE BY SIDE CONVENTIONAL BP SO IT IS EASY TO COMPARE}
    % {THIS IS CONVENTIONAL BP, JUST REWRITTEN in matrix form}
    \var{H = H.unsqueeze(dim=0).T}\\
    \var{C = llr.unsqueeze(dim=-1)}\\
    \For{t in range(iters)}{
    \var{Q = C \textcolor{blue}{if} t == 0 \textcolor{blue}{else} C + sum(R*H,dim=-1).unsqueeze(dim=-1) - R}\\
        \var{tmp = tanh(0.5*Q)}\\
        \var{R = 2*atanh( prod(tmp*H+(1-H),dim=1)/tmp )} \\
        % \Comment{(1-H) for identity element in prod.}
    }
    \Return{C.squeeze()+sum(R*H,dim=-1)}
  }
    \caption{Tensor Belief Propagation}
      \label{algo:bp_tensor}
\end{algorithm}

%%%%%%%%%%%%%%%%%%%%%%%%%%%%%%%%%%%%%%%%%%%%%%%%%%%%%%%%%%%%%%%%%%%%%%%%%%%%%%%%%%%%%%%%%%%%%%%%%%%%%%%%%%%%%%%%%%%%%
%%%%%%%%%%%%%%%%%%%%%%%%%%%%%%%%%%%%%%%%%%%%%%%%%%%%%%%%%%%%%%%%%%%%%%%%%%%%%%%%%%%%%%%%%%%%%%%%%%%%%%%%%%%%%%%%%%%%%
%%%%%%%%%%%%%%%%%%%%%%%%%%%%%%%%%%%%%%%%%%%%%%%%%%%%%%%%%%%%%%%%%%%%%%%%%%%%%%%%%%%%%%%%%%%%%%%%%%%%%%%%%%%%%%%%%%%%%
%%%%%%%%%%%%%%%%%%%%%%%%%%%%%%%%%%%%%%%%%%%%%%%%%%%%%%%%%%%%%%%%%%%%%%%%%%%%%%%%%%%%%%%%%%%%%%%%%%%%%%%%%%%%%%%%%%%%%
%%%%%%%%%%%%%%%%%%%%%%%%%%%%%%%%%%%%%%%%%%%%%%%%%%%%%%%%%%%%%%%%%%%%%%%%%%%%%%%%%%%%%%%%%%%%%%%%%%%%%%%%%%%%%%%%%%%%%
%%%%%%%%%%%%%%%%%%%%%%%%%%%%%%%%%%%%%%%%%%%%%%%%%%%%%%%%%%%%%%%%%%%%%%%%%%%%%%%%%%%%%%%%%%%%%%%%%%%%%%%%%%%%%%%%%%%%%
%%%%%%%%%%%%%%%%%%%%%%%%%%%%%%%%%%%%%%%%%%%%%%%%%%%%%%%%%%%%%%%%%%%%%%%%%%%%%%%%%%%%%%%%%%%%%%%%%%%%%%%%%%%%%%%%%%%%%
%%%%%%%%%%%%%%%%%%%%%%%%%%%%%%%%%%%%%%%%%%%%%%%%%%%%%%%%%%%%%%%%%%%%%%%%%%%%%%%%%%%%%%%%%%%%%%%%%%%%%%%%%%%%%%%%%%%%%
%%%%%%%%%%%%%%%%%%%%%%%%%%%%%%%%%%%%%%%%%%%%%%%%%%%%%%%%%%%%%%%%%%%%%%%%%%%%%%%%%%%%%%%%%%%%%%%%%%%%%%%%%%%%%%%%%%%%%

\paragraph{Belief Propagation Codes Optimization}
The tensor reformulation solves the major challenge of graph learning (challenge (iv)). Challenges (ii) and (iii) are also eliminated in our formulation. 
First, since for any given $H$ the conditional independence of error probability under symmetry \citep{richardson2001capacity} is satisfied for message passing algorithms, it is enough to optimize the zero codeword only, i.e., $c=Gm=0$, removing then any dependency on $G$ in the objective (challenge (ii)). 
As a byproduct, we obtain that the optimization problem is invariant to the choice of modulation, whether differentiable or not (challenge (iii)).

To optimize $H$ (challenge (i)) we relax the NP-hard binary programming problems to an unconstrained objective where, given a parameter matrix $\Omega \in \mathbb{R}^{(n-k)\times n}$, we have \mbox{$H\coloneqq H(\Omega)=\bin(\Omega)$}. Here $\bin(\cdot)$ refers to the element-wise binarization operator implemented via the shifted straight-through-estimator (STE) \citep{bengio2013estimating} defined such that 

%{\color{red}WHY DO YOU GIVE 0 TO POSITIVE?}
%{\color{blue}assume  x binary so: y = 1-2x -> x = 0.5(1-y)}

\begin{equation}
\begin{aligned}
\label{eq:bin_func}
%\begin{cases}
    \bin(u) = (1-\text{sign}(u))/2\,,\quad   {\partial\bin(u)}/{\partial u} \coloneqq -0.5\mathbbm{1}_{|u|\leq 1}
%\end{cases}
\end{aligned}
\end{equation}

Finally, opting for the binary cross-entropy loss (BCE) as the Bit Error rate (BER) discrepancy measure $\mathcal{D}=\text{BCE}$, we obtain the following empirical risk objective
\begin{equation}
\begin{aligned}
\label{eq:e2e_obj_emp}
\mathcal{L}(\Omega) 
% = \frac{1}{nT}
=\sum_{t=1}^{T}\sum_{i=1}^{n} \text{BCE}\bigg(f_{\bin(\Omega),t}\big(c_{s}+{\varepsilon_{i}}\big),c\bigg)+\mathcal{R}(\bin(\Omega))
% \argmin_{H \in \{0,1\}^{(n-k)\times n}} \mathbb{E}_{m\sim \text{Bern}^{k}({\nicefrac{1}{2}}), \varepsilon \sim \mathcal{Z}, T\in \mathbb{N}_{+}} 
\end{aligned}
\end{equation}
where $c_{s}=\phi(c)$ denotes the modulated \emph{zero codeword} and $\varepsilon_{i}$ denotes the $i^{th}$ noise sample drawn from the channel noise distribution. 
This objective aims to provide optimal decoding on different numbers of (variable) decoding iterations $t$. 

While highly non-convex, the objective is (sub)differentiable when considering the STE definition of the gradient \citep{bengio2013estimating, yin2019understanding} and thus optimizable via classical first-order methods.
Since $H$ is binary, only changes in the sign of $\Omega$ are relevant for the optimization, so most gradient descent iterations remain ineffective in reducing the objective using conventional small learning-rate regimes. 
Thus, given the gradient $\nabla_{\Omega}\mathcal{L}$ computed on sufficient statistics, we propose a line-search procedure capable of finding the optimal step size. 

\paragraph{Binary Line-Search} Conventional first-order optimization methods with small learning rate regimes have two major drawbacks with binarization \citep{xnor_net,courbariaux2016binarized}. First, they are generally slow since only gradient steps modifying the sign of the binarized tensor induce a modification of the loss. Second, they have difficulties in converging to local minima because of oscillating behavior around zero.

In general, efficient line search methods \citep{Nocedal2006} assume local convexity or a smooth objective \citep{wolf1978efficient} or, alternatively, apply exhaustive search on a given interval. Since this is not our case, we propose a novel efficient \emph{grid}-search approach optimized to our binary programming setting.
While classical grid search methods look for the optimal step size on handcrafted predefined sample points, in our binary setting we can search only for the step sizes inducing a flip of the sign in $\Omega$, provably limiting the maximum number of \emph{relevant} grid samples to $n(n-k)$.
Thus, the line-search problem is now given by
\begin{equation}
\begin{aligned}
\label{eq:ls_obj}
\lambda^{*} = \argmin_{\lambda \in \mathcal{I}_{\Omega}} 
\mathcal{L}(\Omega-\lambda \nabla_{\Omega}\mathcal{L}) ,\ \ \ \ \ \ \ \mathcal{I}_{\Omega}=\{ s_{i}=\frac{(\Omega)_{i}}{(\nabla_{\Omega}\mathcal{L})_{i}} |s_{i}>0\}\,,
\end{aligned}
\end{equation}
which corresponds to the (parallelizable) objective evaluation on the obtained grid. 
The same formulation can support other more practical line-search objectives instead of the cross-entropy loss $\mathcal{L}$, such as BER or Frame Error Rate (FER). 

% {\color{red}ADD A LINE BY LINE BREAKDOWN. AND EASY WAY TO FILL A PAPER + VEry USEFUL TO THE READERS}
\begin{algorithm}[t]
\footnotesize 
  \Function{Loss(H, x, y, BPiters=5)}{
   \Comment{H is the initial binary parity-check matrix, (n-k,n)}
   \Comment{BPiters is the number of BP iterations}
        \Return{\var{BCE(BP(computeLLR(y), H, BPiters),x)}}
    }   
  \Function{BPCodesOptimization(H, iters)}{
   \Comment{H is the initial binary parity-check matrix, (n-k,n)}
   \Comment{iters is the number of optimization iterations}
   \var{$\Omega$ = 1-2*H}\\
    \For{t in range(iters)}{
        \var{x,y = getData() } \\
        \var{Loss(bin($\Omega$),x,y).backward()}\\
        \var{lambdas = $\Omega$/$\Omega$.grad}\\
        \var{lambdas = sorted(lambdas[lambdas>0].view(-1))[:50]}\\
        \var{$\Omega$ = $\Omega$ - $\Omega$.grad*lambdas[argmin([Loss(bin($\Omega$ -lambda*$\Omega$.grad),x,y) for lambda in lambdas]]}\\
        \var{\textcolor{blue}{if} converged: break}
    }
    \Return{bin($\Omega$)}
  }
    \caption{Belief Propagation Codes Optimization}
      \label{algo:full_algo}
\end{algorithm}
%%%%%%%%%%%%%%%%%%%%%%%%%%%%%%%%%%%%%%%%%%%%%%%%%%%%%%%%%%%%%%%%%%%%%%%%%%%%%%%%%%%%%%%%%%%%%%%%%%%%%%%%%%%%%%%%%%%%%
%%%%%%%%%%%%%%%%%%%%%%%%%%%%%%%%%%%%%%%%%%%%%%%%%%%%%%%%%%%%%%%%%%%%%%%%%%%%%%%%%%%%%%%%%%%%%%%%%%%%%%%%%%%%%%%%%%%%%
%%%%%%%%%%%%%%%%%%%%%%%%%%%%%%%%%%%%%%%%%%%%%%%%%%%%%%%%%%%%%%%%%%%%%%%%%%%%%%%%%%%%%%%%%%%%%%%%%%%%%%%%%%%%%%%%%%%%%
%%%%%%%%%%%%%%%%%%%%%%%%%%%%%%%%%%%%%%%%%%%%%%%%%%%%%%%%%%%%%%%%%%%%%%%%%%%%%%%%%%%%%%%%%%%%%%%%%%%%%%%%%%%%%%%%%%%%%
%%%%%%%%%%%%%%%%%%%%%%%%%%%%%%%%%%%%%%%%%%%%%%%%%%%%%%%%%%%%%%%%%%%%%%%%%%%%%%%%%%%%%%%%%%%%%%%%%%%%%%%%%%%%%%%%%%%%%
%%%%%%%%%%%%%%%%%%%%%%%%%%%%%%%%%%%%%%%%%%%%%%%%%%%%%%%%%%%%%%%%%%%%%%%%%%%%%%%%%%%%%%%%%%%%%%%%%%%%%%%%%%%%%%%%%%%%%
%%%%%%%%%%%%%%%%%%%%%%%%%%%%%%%%%%%%%%%%%%%%%%%%%%%%%%%%%%%%%%%%%%%%%%%%%%%%%%%%%%%%%%%%%%%%%%%%%%%%%%%%%%%%%%%%%%%%%
%%%%%%%%%%%%%%%%%%%%%%%%%%%%%%%%%%%%%%%%%%%%%%%%%%%%%%%%%%%%%%%%%%%%%%%%%%%%%%%%%%%%%%%%%%%%%%%%%%%%%%%%%%%%%%%%%%%%%
%%%%%%%%%%%%%%%%%%%%%%%%%%%%%%%%%%%%%%%%%%%%%%%%%%%%%%%%%%%%%%%%%%%%%%%%%%%%%%%%%%%%%%%%%%%%%%%%%%%%%%%%%%%%%%%%%%%%%

\paragraph{Training}
The optimization parameters are the following: the initial $H$ (i.e., initial $\Omega$), the maximum number of optimization steps (if convergence is not reached), the number and quality of the data samples, the grid search length, and the number of BP iterations.

We assume that an initial $H$ is given by the user as the code to be improved. The number of optimization steps is set to 20 iterations. The training noise is sampled randomly per batch in the $\{3,\dots,7\}$ normalized SNR (i.e. $E_b/N_0$) range but can be modified according to the noise setting of interest. The number of data samples per optimization iteration is set to 4.9M for every code as sufficient gradient estimation, and the data samples are required to have non-zero syndrome. Because of computational constraints, the number of BP iterations during training is fixed and set to 5, while other ranges or values of interest can be used instead.
For faster optimization, the grid search is heuristically restricted to the first 50 smallest step sizes as the optimal step size is generally in the vicinity of the working point (Appendix \ref{app:lso}) . Training and experiments are performed on $8\times 12\mathrm{GB}$ GeForce RTX 2080 Ti GPUs and require 2.96 minutes on average per optimization step.

The full training algorithm (pseudocode) is given in Algorithm \ref{algo:full_algo}.
Given an initial parity check matrix, the algorithm optimizes $H$ iteratively upon convergence.
At each iteration, after computing the gradient on sufficiently large statistics {\color{black}(line 7)}, the line search procedure {\color{black}(line 10)} searches for the optimal step size among those that flip the values of $H$ {\color{black}(line 9)}.

\section{Experiments}
\begin{table}[t]
    \centering
    \caption{
    A comparison of the negative natural logarithm of Bit Error Rate (BER) for several normalized SNR values of our method with classical codes. Higher is better. BP results are provided for 5 iterations in the first row and 15 in the second row. The best results are in bold.
    PEG$X$ means the degree of each node is of $X\%$ under the Progressive Edge Growth construction.}
    \label{tab:main_ber_3_snr}
    % \setlength\tabcolsep{1.5pt}
%    \smallskip
% \vspace{-2mm}
    % \resizebox{0.89985\textwidth}{!}{%
    \resizebox{0.99\textwidth}{!}{%
    \begin{tabular}{@{}lc@{~}c@{~}cc@{~}c@{~}cc@{~}c@{~}cc@{~}c@{~}cc@{~}c@{~}cc@{~}c@{~}c@{}}
    \toprule
        Channel & \multicolumn{6}{c}{AWGN} & \multicolumn{6}{c}{Fading} & \multicolumn{6}{c}{Bursting}\\
  %       \cmidrule(lr){2-4}
  %       \cmidrule(lr){5-13}
		% \cmidrule(lr){14-16}
        Method & \multicolumn{3}{c}{BP} & \multicolumn{3}{c}{Our} & \multicolumn{3}{c}{BP} & \multicolumn{3}{c}{Our} &  \multicolumn{3}{c}{BP} & \multicolumn{3}{c}{Our} \\
  %       \cmidrule(lr){2-4}
  %       \cmidrule(lr){5-7}
  %       \cmidrule(lr){8-10}
  %       \cmidrule(lr){11-13}
		% \cmidrule(lr){14-16}
         $E_{b}/N_{0}$ & 4 & 5 & 6  & 4 & 5 & 6  & 4 & 5 & 6  & 4 & 5 & 6  & 4 & 5 & 6  & 4 & 5 & 6   \\ 
         \midrule                                                                                                                                                                                                                                                                                                                                                                                                                                      
		%%%%%%%%%%%%%%%%%%%
  BCH(63,45)      	\comment{AWGN} \comment{BP} & \makecell{4.06\\4.21} & \makecell{4.91\\5.24} & \makecell{6.04\\6.59} 				\comment{Our} &  \makecell{\bf5.44\\ \bf5.70} & \makecell{\bf6.93\\ \bf7.35} & \makecell{\bf8.60\\ \bf9.16} 					\comment{Fading_1.0} \comment{BP} & \makecell{3.09\\3.13} & \makecell{3.46\\3.55} & \makecell{3.90\\4.04} 			\comment{Our} &  \makecell{\bf3.96\\ \bf4.10} & \makecell{\bf4.58\\ \bf4.80} & \makecell{\bf5.27\\ \bf5.56} 						\comment{Bursting} \comment{BP}  & \makecell{3.60\\3.67} & \makecell{4.32\\4.52} & \makecell{5.19\\5.59} 					\comment{Our} & \makecell{\bf4.05\\ \bf4.21} & \makecell{\bf5.07\\ \bf5.40} & \makecell{\bf6.27\\ \bf6.85} 		\\ \midrule
																																																																																																																																																																			
CCSDS(128,64)   	\comment{AWGN} \comment{BP} & \makecell{6.46\\7.32} & \makecell{9.61\\10.83} & \makecell{13.99\\15.43} 		 		\comment{Our} &  \makecell{\bf7.34\\ \bf8.61} & \makecell{\bf10.48\\ \bf12.26} & \makecell{\bf14.37\\ \bf16.00} 				\comment{Fading_1.0} \comment{BP} & \makecell{5.72\\6.43} & \makecell{7.42\\8.29} & \makecell{9.47\\10.28}			\comment{Our} &  \makecell{\bf6.73\\ \bf8.05} & \makecell{\bf8.45\\ \bf10.07} & \makecell{\bf10.45\\ \bf12.37} 	 					\comment{Bursting} \comment{BP}  & \makecell{5.29\\5.98} & \makecell{7.81\\8.85} & \makecell{11.25\\12.53} 		 			\comment{Our} & \makecell{\bf6.23\\ \bf7.39} & \makecell{\bf8.80\\ \bf10.43} & \makecell{\bf11.90\\ \bf13.28} 	\\ \midrule
																																																																																																																																																																			
LDPC(121,60)    	\comment{AWGN} \comment{BP} & \makecell{4.81\\5.31} & \makecell{7.17\\7.96} & \makecell{10.75\\11.85} 				\comment{Our} &  \makecell{\bf7.70\\ \bf8.86} & \makecell{\bf10.87\\ \bf11.91} & \makecell{\bf14.25\\ \bf14.41} 				\comment{Fading_1.0} \comment{BP} & \makecell{4.10\\4.42} & \makecell{5.23\\5.61} & \makecell{6.68\\7.04} 			\comment{Our} &  \makecell{\bf6.68\\ \bf7.71} & \makecell{\bf8.47\\ \bf9.67} & \makecell{\bf10.50\\ \bf11.76} 	 					\comment{Bursting} \comment{BP}  & \makecell{3.97\\4.31} & \makecell{5.75\\6.37} & \makecell{8.40\\9.25} 					\comment{Our} & \makecell{\bf6.23\\ \bf7.26} & \makecell{\bf8.89\\ \bf10.03} & \makecell{\bf11.98\\ \bf12.88} 	\\ \midrule
																																																																																																																																																																			
LDPC(121,80)    	\comment{AWGN} \comment{BP} & \makecell{6.59\\7.35} & \makecell{9.68\\10.94} & \makecell{13.43\\15.46} 		 		\comment{Our} &  \makecell{\bf7.77\\ \bf8.75} & \makecell{\bf11.21\\ \bf12.45} & \makecell{\bf15.06\\ \bf15.67} 				\comment{Fading_1.0} \comment{BP} & \makecell{4.60\\4.97} & \makecell{5.80\\6.29} & \makecell{7.22\\7.82} 			\comment{Our} &  \makecell{\bf5.55\\ \bf6.25} & \makecell{\bf6.90\\ \bf7.80} & \makecell{\bf8.36\\ \bf9.47} 	 					\comment{Bursting} \comment{BP}  & \makecell{5.30\\5.81} & \makecell{7.60\\8.50} & \makecell{10.66\\12.15} 					\comment{Our} & \makecell{\bf6.23\\ \bf6.99} & \makecell{\bf8.87\\ \bf10.09} & \makecell{\bf12.19\\ \bf13.74} 	\\ \midrule
																																																																																																																																																																			
LDPC(128,64)    	\comment{AWGN} \comment{BP} & \makecell{3.66\\4.00} & \makecell{4.65\\5.16} & \makecell{5.80\\6.42} 				\comment{Our} &  \makecell{\bf5.54\\ \bf6.56} & \makecell{\bf7.37\\ \bf8.70} & \makecell{\bf9.44\\ \bf10.81} 					\comment{Fading_1.0} \comment{BP} & \makecell{3.22\\3.51} & \makecell{3.80\\4.18} & \makecell{4.44\\4.84} 			\comment{Our} &  \makecell{\bf4.86\\ \bf5.64} & \makecell{\bf5.94\\ \bf6.85} & \makecell{\bf7.15\\ \bf8.14} 						\comment{Bursting} \comment{BP}  & \makecell{3.23\\3.48} & \makecell{4.08\\4.51} & \makecell{5.09\\5.66} 					\comment{Our} & \makecell{\bf3.72\\ \bf4.13} & \makecell{\bf5.00\\ \bf5.72} & \makecell{\bf6.54\\ \bf7.66} 		\\ \midrule
																																																																																																																																																																			
LDPC(32,16)     	\comment{AWGN} \comment{BP} & \makecell{4.36\\4.64} & \makecell{5.59\\6.07} & \makecell{7.18\\7.94} 				\comment{Our} &  \makecell{\bf5.48\\ \bf5.76} & \makecell{\bf7.02\\ \bf7.44} & \makecell{\bf8.92\\ \bf9.41} 					\comment{Fading_1.0} \comment{BP} & \makecell{4.03\\4.29} & \makecell{4.70\\5.06} & \makecell{5.47\\5.90} 			\comment{Our} &  \makecell{\bf5.26\\ \bf5.43} & \makecell{\bf6.02\\ \bf6.23} & \makecell{\bf6.82\\ \bf6.97} 						\comment{Bursting} \comment{BP}  & \makecell{3.88\\4.09} & \makecell{4.89\\5.26} & \makecell{6.18\\6.76} 					\comment{Our} & \makecell{\bf4.77\\ \bf5.01} & \makecell{\bf6.02\\ \bf6.35} & \makecell{\bf7.52\\ \bf7.96} 		\\ \midrule
																																																																																																																																																																			
LDPC(96,48)     	\comment{AWGN} \comment{BP} & \makecell{6.73\\7.50} & \makecell{9.48\\10.61} & \makecell{12.98\\ \bf14.26} 	 		\comment{Our} &  \makecell{\bf7.22\\ \bf8.29} & \makecell{\bf9.96\\ \bf11.12} & \makecell{\bf13.37\\14.06} 						\comment{Fading_1.0} \comment{BP} & \makecell{3.83\\4.17} & \makecell{4.57\\4.94} & \makecell{5.35\\5.73} 			\comment{Our} &  \makecell{\bf5.37\\ \bf6.14} & \makecell{\bf6.51\\ \bf7.38} & \makecell{\bf7.71\\ \bf8.65} 						\comment{Bursting} \comment{BP}  & \makecell{5.68\\6.33} & \makecell{7.94\\8.91} & \makecell{10.90\\ \bf11.99} 	 			\comment{Our} & \makecell{\bf5.90\\ \bf6.71} & \makecell{\bf8.19\\ \bf9.28} & \makecell{\bf10.91\\11.75} 		\\ \midrule
																																																																																																																																																																			
LTE(132,40)     	\comment{AWGN} \comment{BP} & \makecell{2.94\\3.37} & \makecell{3.32\\3.79} & \makecell{3.57\\4.09} 				\comment{Our} &  \makecell{\bf3.25\\ \bf3.93} & \makecell{\bf3.71\\ \bf4.49} & \makecell{\bf4.04\\ \bf4.89} 					\comment{Fading_1.0} \comment{BP} & \makecell{3.17\\3.60} & \makecell{3.45\\3.82} & \makecell{3.67\\4.01} 			\comment{Our} &  \makecell{\bf4.49\\ \bf5.32} & \makecell{\bf4.99\\ \bf5.81} & \makecell{\bf5.47\\ \bf6.31} 						\comment{Bursting} \comment{BP}  & \makecell{2.75\\3.17} & \makecell{3.17\\3.62} & \makecell{3.47\\3.96} 					\comment{Our} & \makecell{\bf2.99\\ \bf3.53} & \makecell{\bf3.44\\ \bf4.03} & \makecell{\bf3.78\\ \bf4.41} 		\\ \midrule
																																																																																																																																																																			
MACKAY(96,48)   	\comment{AWGN} \comment{BP} & \makecell{6.75\\7.59} & \makecell{9.45\\10.52} & \makecell{\bf12.85\\ \bf14.09}		\comment{Our} &  \makecell{\bf7.03\\ \bf7.99} & \makecell{\bf9.63\\ \bf10.97} & \makecell{12.78\\14.05} 						\comment{Fading_1.0} \comment{BP} & \makecell{6.28\\7.04} & \makecell{7.86\\8.76} & \makecell{9.55\\10.64} 			\comment{Our} &  \makecell{\bf6.53\\ \bf7.47} & \makecell{\bf8.06\\ \bf9.32} & \makecell{\bf9.77\\ \bf11.19} 	 					\comment{Bursting} \comment{BP}  & \makecell{5.72\\6.39} & \makecell{7.97\\8.90} & \makecell{10.81\\11.91} 					\comment{Our} & \makecell{\bf5.95\\ \bf6.82} & \makecell{\bf8.23\\ \bf9.41} & \makecell{\bf10.91\\ \bf12.71} 	\\ \midrule
																																																																																																																																																																			
POLAR(128,86)   	\comment{AWGN} \comment{BP} & \makecell{3.76\\4.02} & \makecell{4.17\\4.67} & \makecell{4.58\\5.38} 				\comment{Our} &  \makecell{\bf4.83\\ \bf5.37} & \makecell{\bf5.87\\ \bf6.88} & \makecell{\bf6.58\\ \bf8.10} 					\comment{Fading_1.0} \comment{BP} & \makecell{3.15\\3.28} & \makecell{3.53\\3.73} & \makecell{3.91\\4.18} 			\comment{Our} &  \makecell{\bf3.64\\ \bf3.92} & \makecell{\bf4.28\\ \bf4.70} & \makecell{\bf4.94\\ \bf5.52} 						\comment{Bursting} \comment{BP}  & \makecell{3.48\\3.65} & \makecell{3.96\\4.31} & \makecell{4.37\\4.97} 					\comment{Our} & \makecell{\bf3.69\\ \bf3.87} & \makecell{\bf4.51\\ \bf4.91} & \makecell{\bf5.18\\ \bf5.91} 		\\ \midrule
																																																																																																																																																																			
RS(60,52)       	\comment{AWGN} \comment{BP} & \makecell{4.41\\4.54} & \makecell{5.32\\5.52} & \makecell{6.41\\6.64} 				\comment{Our} &  \makecell{\bf5.02\\ \bf5.07} & \makecell{\bf6.38\\ \bf6.47} & \makecell{\bf7.99\\ \bf8.12} 					\comment{Fading_1.0} \comment{BP} & \makecell{3.11\\3.13} & \makecell{3.41\\3.43} & \makecell{3.77\\3.81} 			\comment{Our} &  \makecell{\bf3.37\\ \bf3.38} & \makecell{\bf3.73\\ \bf3.75} & \makecell{\bf4.12\\ \bf4.15} 						\comment{Bursting} \comment{BP}  & \makecell{3.85\\3.91} & \makecell{4.58\\4.72} & \makecell{5.44\\5.67} 					\comment{Our} & \makecell{\bf4.17\\ \bf4.21} & \makecell{\bf5.18\\ \bf5.27} & \makecell{\bf6.40\\ \bf6.56} 		\\  \midrule

LDPC PEG2(64,32)     \comment{AWGN} \comment{BP} & \makecell{4.38\\4.38} & \makecell{5.12\\5.13} & \makecell{6.04\\6.04} \comment{Our} & \makecell{\bf4.45\\ \bf4.44} & \makecell{\bf5.19\\ \bf5.19} & \makecell{\bf6.10\\ \bf6.10} \comment{Fading_1.0} \comment{BP} & \makecell{4.08\\4.08} & \makecell{4.44\\4.44} & \makecell{4.81\\4.81} \comment{Our} & \makecell{\bf4.10\\ \bf4.10} & \makecell{\bf4.46\\ \bf4.47} & \makecell{\bf4.85\\ \bf4.85} \comment{Bursting} \comment{BP} & \makecell{4.07\\4.06} & \makecell{4.69\\4.69} & \makecell{5.43\\5.43} \comment{Our} & \makecell{\bf4.07\\ \bf4.06} & \makecell{\bf4.70\\ \bf4.69} & \makecell{\bf5.43\\ \bf5.44} \\ \midrule

LDPC PEG5(64,32)     \comment{AWGN} \comment{BP} & \makecell{6.02\\6.63} & \makecell{8.20\\9.06} & \makecell{10.95\\ \bf12.30} \comment{Our} & \makecell{\bf6.53\\ \bf7.13} & \makecell{\bf8.73\\ \bf9.48} & \makecell{\bf11.56\\12.20} \comment{Fading_1.0} \comment{BP} & \makecell{5.63\\6.19} & \makecell{6.86\\7.52} & \makecell{8.31\\9.02} \comment{Our} & \makecell{\bf6.22\\ \bf6.96} & \makecell{\bf7.48\\ \bf8.34} & \makecell{\bf8.82\\ \bf9.85} \comment{Bursting} \comment{BP} & \makecell{5.18\\5.68} & \makecell{6.97\\7.75} & \makecell{9.34\\ \bf10.19} \comment{Our} & \makecell{\bf5.59\\ \bf6.12} & \makecell{\bf7.41\\ \bf8.06} & \makecell{\bf9.51\\10.13} \\ \midrule

LDPC PEG10(64,32)    \comment{AWGN} \comment{BP} & \makecell{3.98\\4.27} & \makecell{5.17\\5.77} & \makecell{6.70\\7.67} \comment{Our} & \makecell{\bf5.56\\ \bf6.25} & \makecell{\bf7.22\\ \bf8.28} & \makecell{\bf9.13\\ \bf10.59} \comment{Fading_1.0} \comment{BP} & \makecell{3.52\\3.71} & \makecell{4.18\\4.47} & \makecell{4.95\\5.30} \comment{Our} & \makecell{\bf5.02\\ \bf5.60} & \makecell{\bf6.00\\ \bf6.72} & \makecell{\bf7.11\\ \bf7.90} \comment{Bursting} \comment{BP} & \makecell{3.48\\3.67} & \makecell{4.47\\4.90} & \makecell{5.75\\6.46} \comment{Our} & \makecell{\bf4.26\\ \bf4.73} & \makecell{\bf5.50\\ \bf6.22} & \makecell{\bf7.01\\ \bf8.01} \\

\bottomrule
	\end{tabular}
	}
\end{table} 

We evaluate our framework on five classes of linear codes: various Low-Density Parity Check (LDPC) codes \citep{gallager1962low, abu2010trapping}, Polar codes \citep{arikan2008channel}, Reed Solomon codes \citep{reed1960polynomial}, Bose–Chaudhuri–Hocquenghem (BCH) codes \citep{bose1960class} and random codes. 
%{\color{red}PROBLEMATIC STATEMENT However, we focus the experiments on the more interesting sparse codes since LDPC codes are optimized for BP decoding \citep{richardson2001design}.}
%{\color{blue}WHY PROBLEMATIC? SHOWING YOU IMPROVE BCH IS NOT SO INTERESTING. IF YOU CAN IMPROVE OVER LDPC CODES THEN YOU HAVE A STRONG CLAIM. OTHER CODES ARE JUST HERE TO SHOW I AM TRULY IMPROVING (SANITY CHECK). }
All the parity check matrices are taken from \citep{channelcodes} except the LDPC codes created using the popular Progressive Edge Growth framework \citep{hu2005regular,pge_imp}. 

We consider three types of channel noise under BPSK modulation. We first test our framework with the canonical AWGN channel given as $y = c_{s}+\varepsilon$ with $\varepsilon\sim \mathcal{N}(0,\sigma I_{n})$.
We also consider the Rayleigh fading channel, where $y = h\odot c_{s}+\varepsilon$, with $h$ the iid Rayleigh distributed fading vector with coefficient $1$ and $\varepsilon$ the regular AWGN noise, where we assume ideal channel state information. Finally, we consider the AWGN channel with Gaussian mixture channel (also referred to as bursty noise) simulating wireless channel interference as $y = c_{s}+\varepsilon+\zeta$ with $\varepsilon$ the AWGN and $\zeta_{i}\sim \mathcal{N}(0,\sqrt{2}\sigma)$ with probability $\rho=0.1$ and $\zeta_{i}=0$ with probability $1-\rho$.

The results are reported as negative natural logarithm bit error rates (BER) for three different normalized SNR values ($E_{b}/N_{0}$), following the conventional testing benchmark, e.g., \citep{nachmani2019hyper,choukroun2022error}. BP-based results are obtained after $\ell=5$ BP iterations in the first row and $\ell=15$ in the second row of the results tables. During testing, at least $10^{5}$ random codewords are decoded, to obtain at least $50$ frames with errors at each SNR value.
 For this section, we performed a small hyperparameter search as reported in Appendix \ref{app:hyperparam}, where the final code is selected to have the lowest average BER on the SNR test range.

The results are provided in Table \ref{tab:main_ber_3_snr}, where \emph{our} method means BP applied on the learned code initialized by the given classical code. 
{\color{black}
We also provide in Appendix \ref{app:more_snr} the same table with a broader SNR range ($E_{b}/N_{0} \in \{3,\dots, 7\})$.
}
We provide the improvement statistics (i.e., mean, std, min, max) in dB on all the sparse codes in Figure \ref{fig:stats_sparse} and we extend the analysis to all the codes in Appendix \ref{app:stats_all_codes}. 
For completeness, we provide in Appendix \ref{app:other_baselines} a comparison with the genetic algorithm of \cite{elkelesh2019decoder} where our method demonstrates much better performance while being faster by orders of magnitude.

Evidently, our method improves by large margins all code families on the three different channel noise scenarios and with both numbers of decoding iterations, demonstrating the capacity of the framework to provide improved codes on multiple settings of interest.

\begin{figure}[t]
% \vspace{1em}
\centering
\noindent  \begin{tabular}{@{}ccc@{}}
        \includegraphics[trim={0 0 0 0},clip, width=0.33\linewidth]{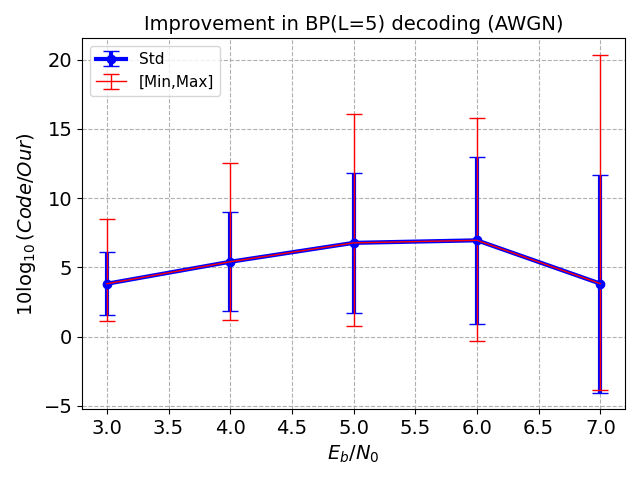}&
        \includegraphics[trim={0 0 0 0},clip, width=0.33\linewidth]{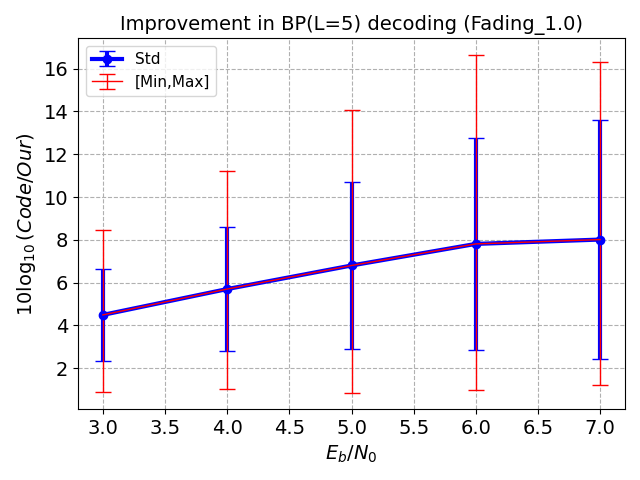}&
        \includegraphics[trim={0 0 0 0},clip, width=0.33\linewidth]{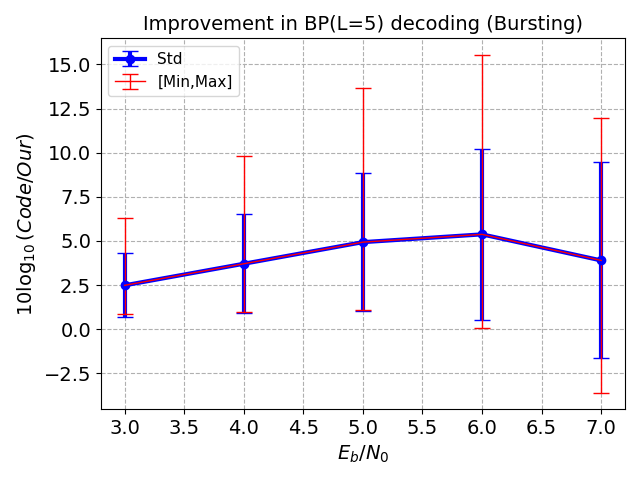}\\
        \includegraphics[trim={0 0 0 0},clip, width=0.33\linewidth]{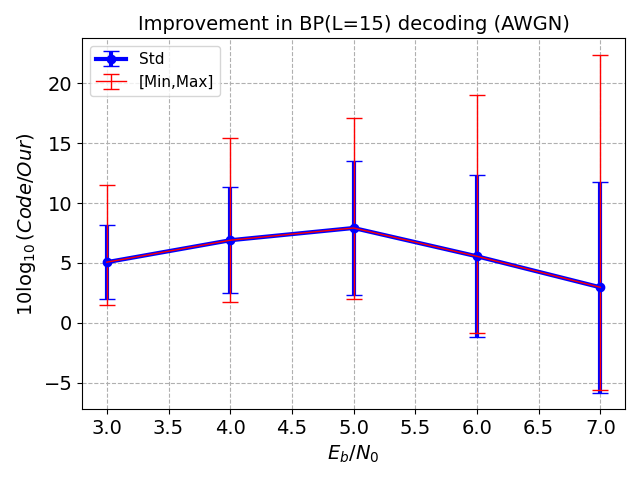}&
        \includegraphics[trim={0 0 0 0},clip, width=0.33\linewidth]{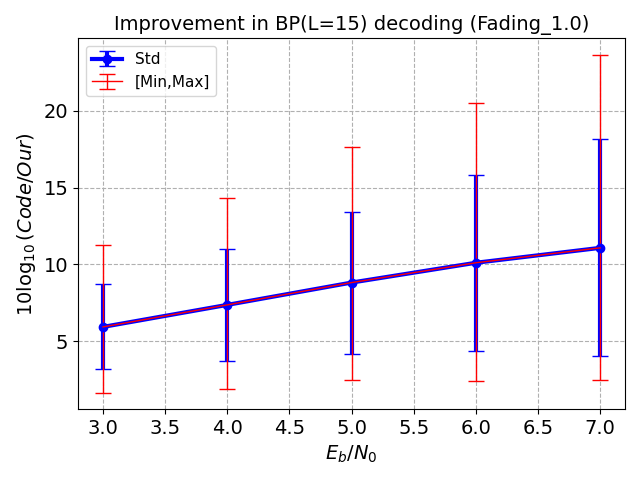}&
        \includegraphics[trim={0 0 0 0},clip, width=0.33\linewidth]{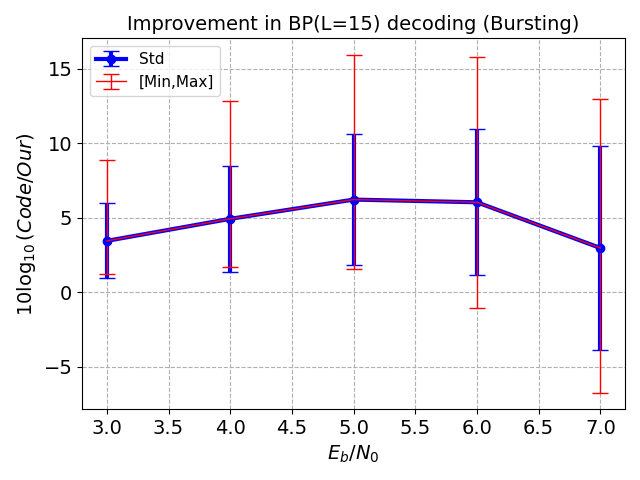}\\
      (a) & (b) & (c)\\
      \end{tabular}
%   \caption{Comparison of the self-attention mechanisms}
  \caption{Statistics of improvement in dB for the  (a) AWGN, (b) fading, and (c) bursting channel on the \emph{sparse codes} only. 
  We provide the mean and standard deviation as well as the minimum and maximum improvements.}
\label{fig:stats_sparse}
\end{figure}

\section{Analysis}
% \paragraph{Large Codes}

\begin{figure}[t]
% \vspace{1em}
\centering
\noindent  \begin{tabular}{@{}cccc@{}}
        \includegraphics[trim={10 0 10 0},clip, width=0.245\linewidth]{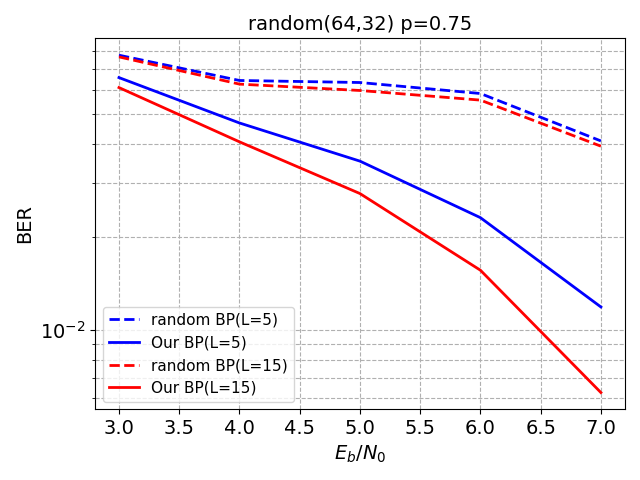}&
        \includegraphics[trim={10 0 10 0},clip, width=0.245\linewidth]{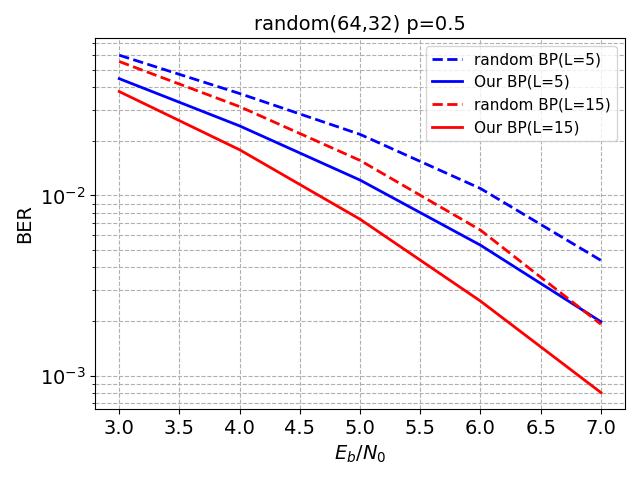}&
        \includegraphics[trim={10 0 10 0},clip, width=0.245\linewidth]{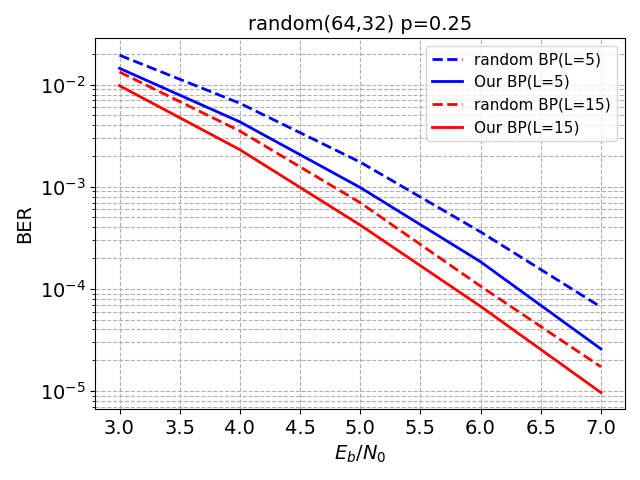}&
        \includegraphics[trim={10 0 10 0},clip, width=0.245\linewidth]{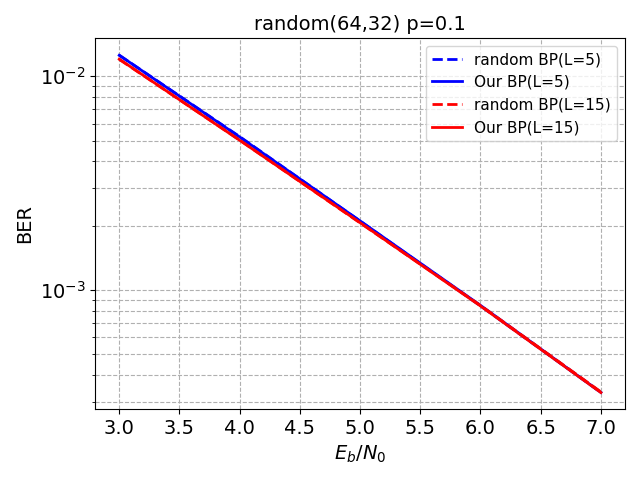}\\
      \end{tabular}
%   \caption{Comparison of the self-attention mechanisms}
  \caption{Performance of the method on random codes under different sparsity rate initialization $p$.}
\label{fig:random_codes_reg}
\end{figure}

\paragraph{Initialization and Random Codes}
\label{subsec:init_random}
We provide in Figure \ref{fig:random_codes_reg} the performance of the proposed method on random codes initialized with different sparsity rates. The parity check matrix is initialized in a systematic form $H=[I_{n-k},P]$ for full rank initialization, where $P\sim \text{Bern}^{(n-k)\times k}(p)$.
We can observe that the framework can greatly improve the performance of the original random code. Most importantly, we can observe that different initializations provide convergence to different local optima and that better initialization generally induces convergence to a better minimum. 
{\color{black}Performance on other code lengths is provided in Appendix \ref{app:more_rand_codes}.}
Finally, good initialization (i.e., (large) well-performing sparse codes under BP decoding) requires perturbation at initialization or during training (c.f., Appendix \ref{app:hyperparam}) in order to get extracted from local minima. 

%%%%%%%%%%%%%%%%%%%%%%%%%%%%%%%%%%%%%%%%%%%%%%%%%%%%%%%%%%%%%%%%%%%%%%%%%%%%%%%%%%%%%%%%%%%%%%%%%%%%%%%%%%%%%%%%%%%%%
%%%%%%%%%%%%%%%%%%%%%%%%%%%%%%%%%%%%%%%%%%%%%%%%%%%%%%%%%%%%%%%%%%%%%%%%%%%%%%%%%%%%%%%%%%%%%%%%%%%%%%%%%%%%%%%%%%%%%
%%%%%%%%%%%%%%%%%%%%%%%%%%%%%%%%%%%%%%%%%%%%%%%%%%%%%%%%%%%%%%%%%%%%%%%%%%%%%%%%%%%%%%%%%%%%%%%%%%%%%%%%%%%%%%%%%%%%%
%%%%%%%%%%%%%%%%%%%%%%%%%%%%%%%%%%%%%%%%%%%%%%%%%%%%%%%%%%%%%%%%%%%%%%%%%%%%%%%%%%%%%%%%%%%%%%%%%%%%%%%%%%%%%%%%%%%%%
%%%%%%%%%%%%%%%%%%%%%%%%%%%%%%%%%%%%%%%%%%%%%%%%%%%%%%%%%%%%%%%%%%%%%%%%%%%%%%%%%%%%%%%%%%%%%%%%%%%%%%%%%%%%%%%%%%%%%
%%%%%%%%%%%%%%%%%%%%%%%%%%%%%%%%%%%%%%%%%%%%%%%%%%%%%%%%%%%%%%%%%%%%%%%%%%%%%%%%%%%%%%%%%%%%%%%%%%%%%%%%%%%%%%%%%%%%%
%%%%%%%%%%%%%%%%%%%%%%%%%%%%%%%%%%%%%%%%%%%%%%%%%%%%%%%%%%%%%%%%%%%%%%%%%%%%%%%%%%%%%%%%%%%%%%%%%%%%%%%%%%%%%%%%%%%%%
%%%%%%%%%%%%%%%%%%%%%%%%%%%%%%%%%%%%%%%%%%%%%%%%%%%%%%%%%%%%%%%%%%%%%%%%%%%%%%%%%%%%%%%%%%%%%%%%%%%%%%%%%%%%%%%%%%%%%
%%%%%%%%%%%%%%%%%%%%%%%%%%%%%%%%%%%%%%%%%%%%%%%%%%%%%%%%%%%%%%%%%%%%%%%%%%%%%%%%%%%%%%%%%%%%%%%%%%%%%%%%%%%%%%%%%%%%%

\begin{figure}[t]
% \vspace{1em}
\centering
\noindent  \begin{tabular}{@{}cccc@{}}
        \includegraphics[trim={10 0 10 0},clip, width=0.245\linewidth]{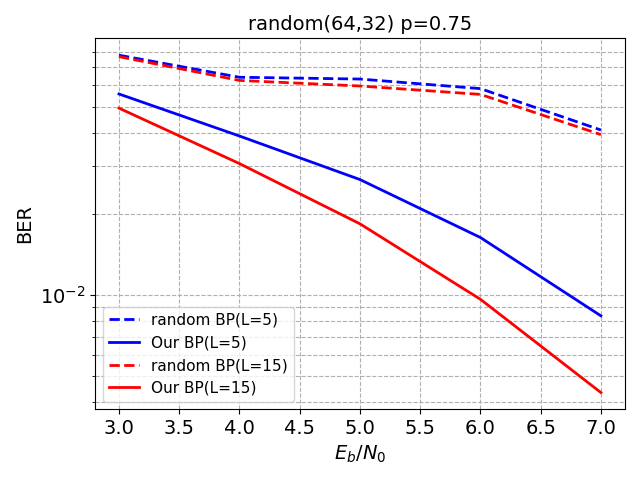}&
        \includegraphics[trim={10 0 10 0},clip, width=0.245\linewidth]{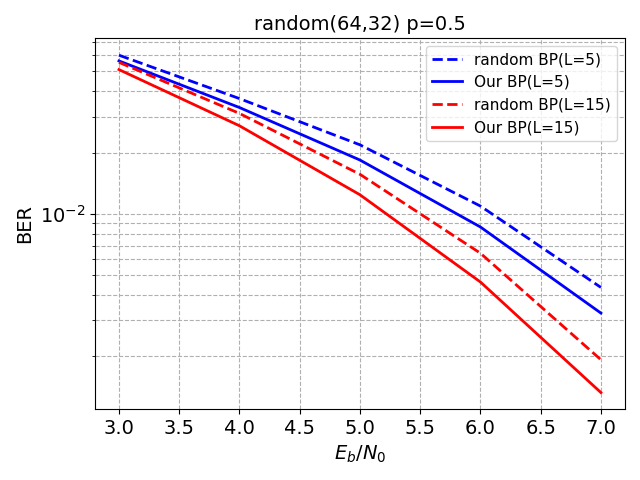}&
        \includegraphics[trim={10 0 10 0},clip, width=0.245\linewidth]{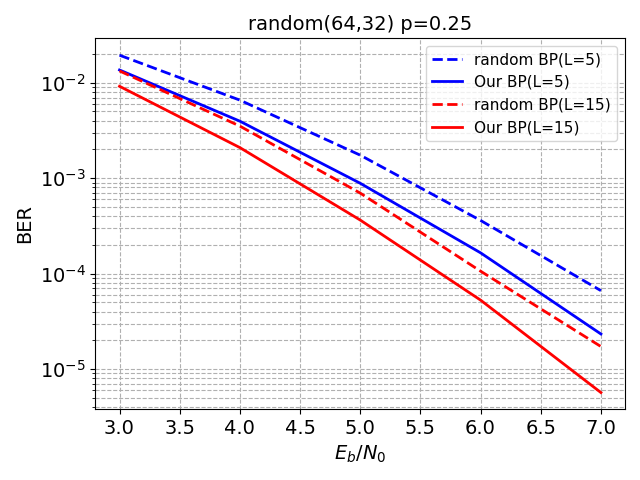}&
        \includegraphics[trim={10 0 10 0},clip, width=0.245\linewidth]{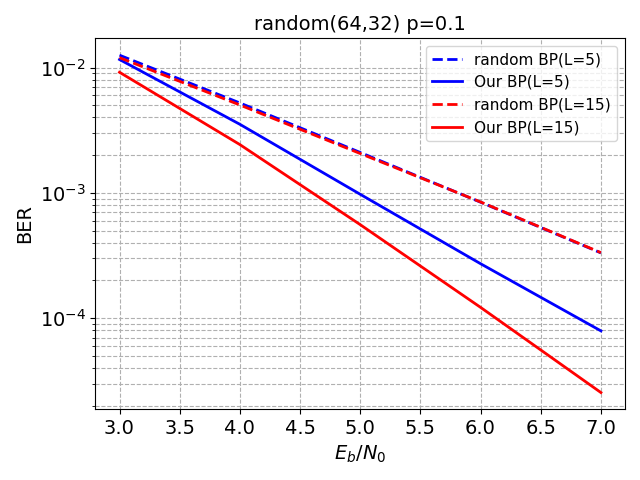}\\
      \end{tabular}
%   \caption{Comparison of the self-attention mechanisms}
  \caption{Performance of the method on constrained systematic random codes under different sparsity rate initialization $p$ on the AWGN channel.}
\label{fig:random_codes_std}
\end{figure}

\paragraph{Constrained Codes}
In Figure \ref{fig:random_codes_std} we provide the performance of the method on constrained systematic random codes as described in the previous paragraph, while here, we constrain them to maintain their systematic form during the optimization, i.e., only the parity matrix elements of $P$ are optimized (via $\Omega$). The optimization is performed by backpropagating over the $P$ tensor only, similarly to having a hard structure constraint on the identity part of $H$. 
While maintaining a structure of interest, we can observe this regularization can further improve the convergence quality (e.g., $p=0.1$) 
%{\color{red}THIS IS JUST ONE RANDOM EXAMPLE PER p} 
compared to the unconstrained setting of Figure \ref{fig:random_codes_reg}.
{\color{black}Performance on other code lengths is provided in Appendix \ref{app:more_rand_codes}.}

\begin{figure}
\centering
\begin{minipage}{.33\textwidth}
\centering
\noindent 
\includegraphics[width=1\textwidth]{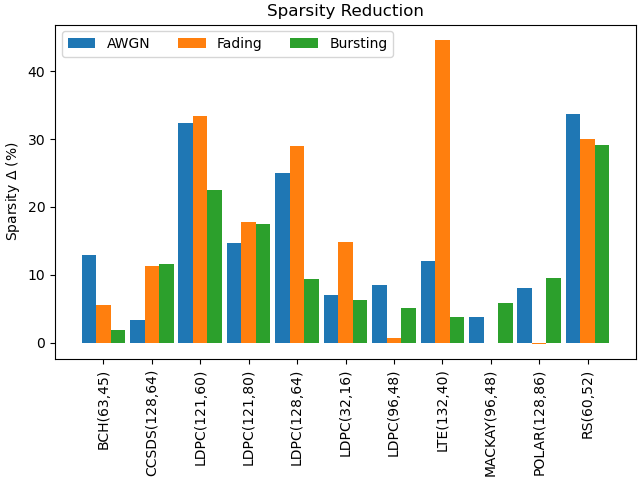}
\caption{Sparsity reduction of the proposed codes.}
\label{fig:sparsity_comp}

\end{minipage}%
\ \ 
\begin{minipage}{.65\textwidth}
\centering
\begin{tabular}{@{}cc@{}}
        \includegraphics[trim={10 0 10 0},clip, width=0.45\linewidth]{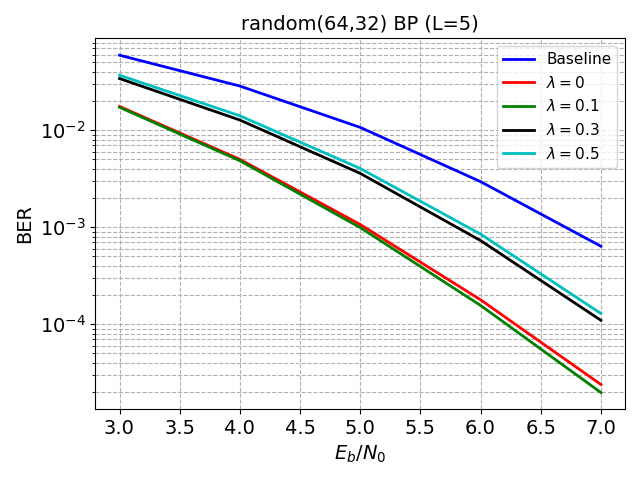}&
        \includegraphics[trim={10 0 10 0},clip, width=0.45\linewidth]{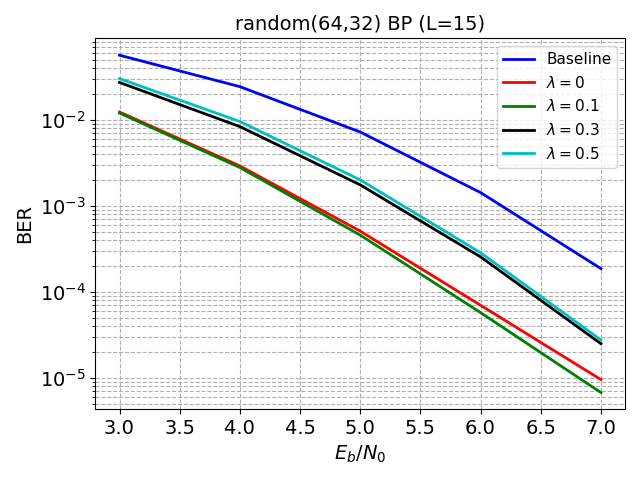}
      \end{tabular}
\caption{Performance of the method with $L_{1}$ regularization for different values of the regularization factor $\lambda$ for random codes with $p=0.25$ on the AWGN channel.
}
\label{fig:random_codes_l1}
\end{minipage}
\end{figure}

In Figure \ref{fig:sparsity_comp} we present the sparsification of the codes created by the framework.
Here, $\Delta=100({S_{b}-S_{o}})/{S_{b}}$ represents the sparsity ratio, with $S_{b}$ and $S_{o}$ being the sparsity of the baseline code and our code, respectively.
We can observe that optimization always provides sparser code. 
Nevertheless, we also observed that the optimization does not modify the girth of the code.

In Figure \ref{fig:random_codes_l1} we present the performance of the method on random codes with sparsity constraint, i.e., $\mathcal{R}(H)=\lambda \|H\|_{1}$, with $\lambda \in \mathbb{R}_{+}$.
We can observe that adding a sparsity constraint is generally not profitable, since the optimization over BP already induces sparse codes.

\begin{figure}[t]
% \vspace{1em}
\centering
\noindent  \begin{tabular}{@{}cccc@{}}
        \includegraphics[trim={10 0 10 0},clip, width=0.23\linewidth]{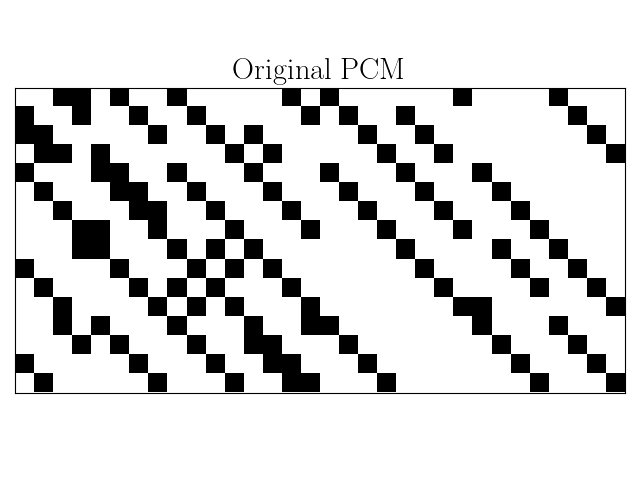}&
        \includegraphics[trim={10 0 10 0},clip, width=0.23\linewidth]{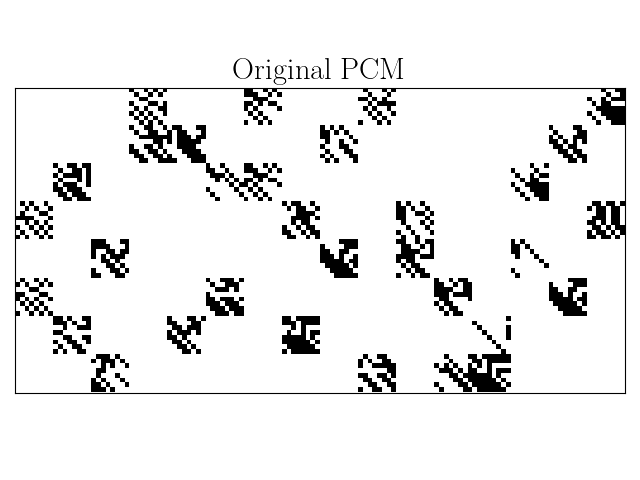}&
        \includegraphics[trim={10 0 10 0},clip, width=0.23\linewidth]{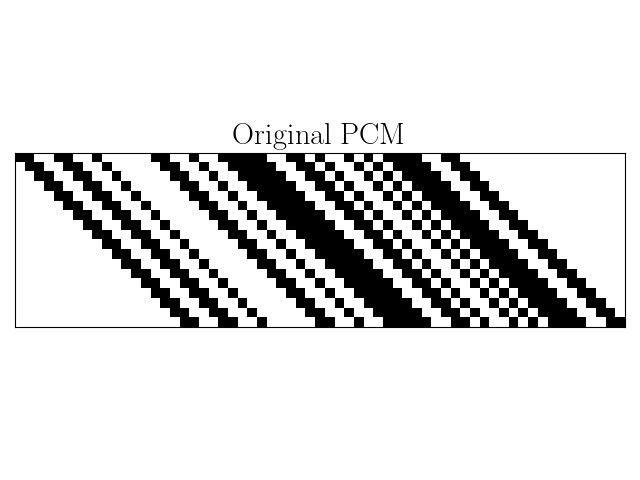}&
        \includegraphics[trim={10 0 10 0},clip, width=0.23\linewidth]{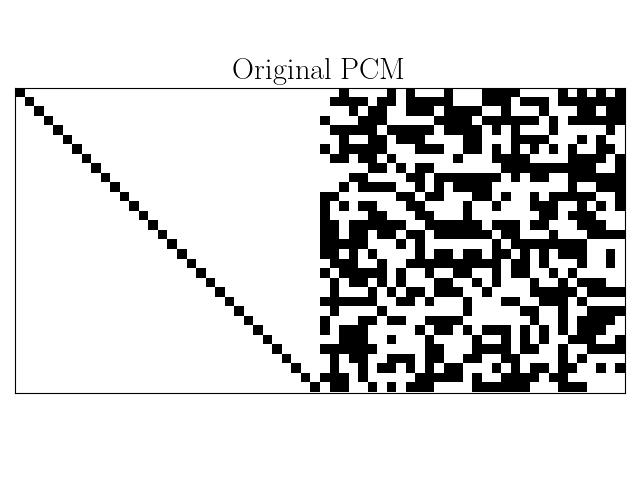}\\
        %%%%
        \includegraphics[trim={10 0 10 0},clip, width=0.23\linewidth]{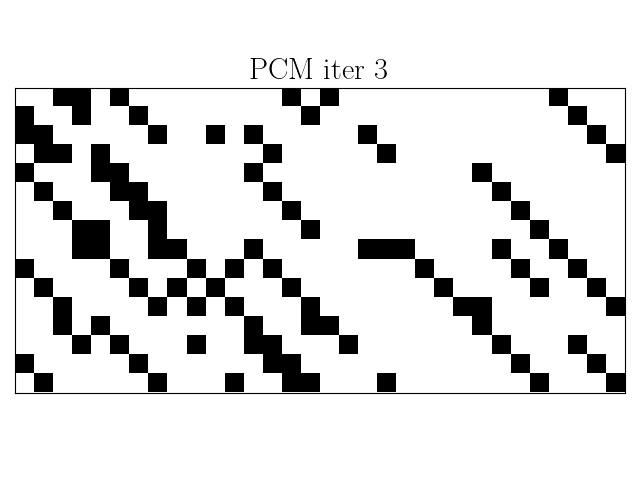}&
        \includegraphics[trim={10 0 10 0},clip, width=0.23\linewidth]{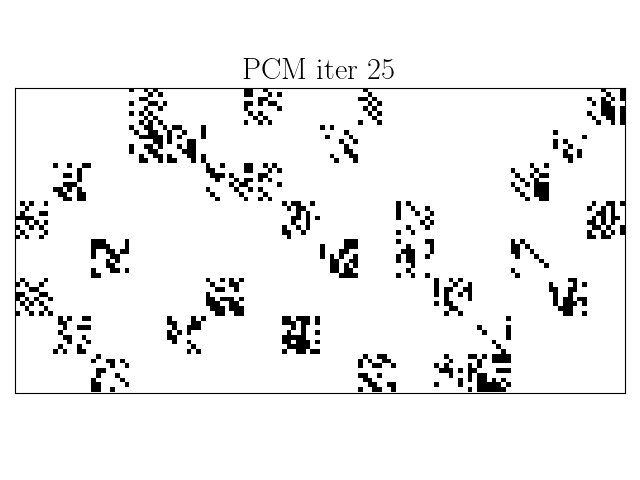}&
        \includegraphics[trim={10 0 10 0},clip, width=0.23\linewidth]{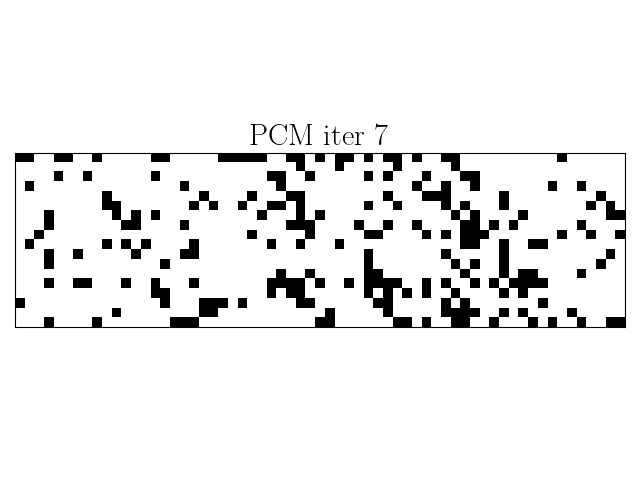}&
        \includegraphics[trim={10 0 10 0},clip, width=0.23\linewidth]{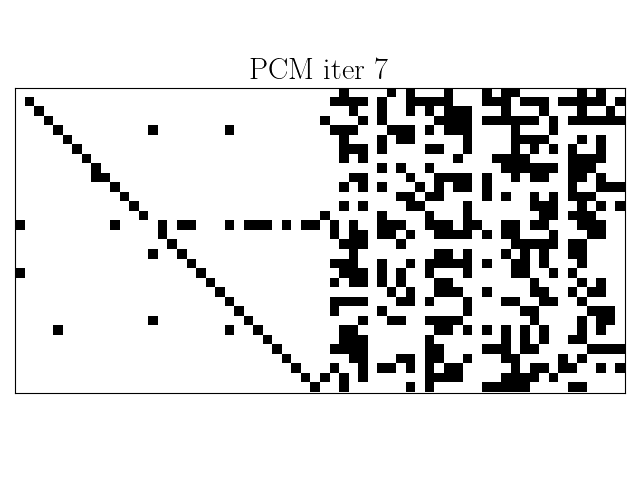}\\
        (a) & (b) & (c) & (d)
      \end{tabular}
%   \caption{Comparison of the self-attention mechanisms}
  \caption{Visualization of the original (first row) and the learned parity check matrices (second row) for (a) LDPC(32,16), (b) LDPC(128,64), (c) BCH(63,45) and (d) Random(64,32,$p=0.5$). "PCM iter X" denotes the final iteration Parity Check Matrix of the optimization.}
\label{fig:pc_mats}
\end{figure}

%%%%%%%%%%%%%%%%%%%%%%%%%%%%%%%%%%%%%%%%%%%%%%%%%%%%%%%%%%%%%%%%%%%%%%%%%%%%%%%%%%%%%%%%%%%%%%%%%%%%%%%%%%%%%%%%%%%%%
%%%%%%%%%%%%%%%%%%%%%%%%%%%%%%%%%%%%%%%%%%%%%%%%%%%%%%%%%%%%%%%%%%%%%%%%%%%%%%%%%%%%%%%%%%%%%%%%%%%%%%%%%%%%%%%%%%%%%
%%%%%%%%%%%%%%%%%%%%%%%%%%%%%%%%%%%%%%%%%%%%%%%%%%%%%%%%%%%%%%%%%%%%%%%%%%%%%%%%%%%%%%%%%%%%%%%%%%%%%%%%%%%%%%%%%%%%%
%%%%%%%%%%%%%%%%%%%%%%%%%%%%%%%%%%%%%%%%%%%%%%%%%%%%%%%%%%%%%%%%%%%%%%%%%%%%%%%%%%%%%%%%%%%%%%%%%%%%%%%%%%%%%%%%%%%%%
%%%%%%%%%%%%%%%%%%%%%%%%%%%%%%%%%%%%%%%%%%%%%%%%%%%%%%%%%%%%%%%%%%%%%%%%%%%%%%%%%%%%%%%%%%%%%%%%%%%%%%%%%%%%%%%%%%%%%
%%%%%%%%%%%%%%%%%%%%%%%%%%%%%%%%%%%%%%%%%%%%%%%%%%%%%%%%%%%%%%%%%%%%%%%%%%%%%%%%%%%%%%%%%%%%%%%%%%%%%%%%%%%%%%%%%%%%%
%%%%%%%%%%%%%%%%%%%%%%%%%%%%%%%%%%%%%%%%%%%%%%%%%%%%%%%%%%%%%%%%%%%%%%%%%%%%%%%%%%%%%%%%%%%%%%%%%%%%%%%%%%%%%%%%%%%%%
%%%%%%%%%%%%%%%%%%%%%%%%%%%%%%%%%%%%%%%%%%%%%%%%%%%%%%%%%%%%%%%%%%%%%%%%%%%%%%%%%%%%%%%%%%%%%%%%%%%%%%%%%%%%%%%%%%%%%
%%%%%%%%%%%%%%%%%%%%%%%%%%%%%%%%%%%%%%%%%%%%%%%%%%%%%%%%%%%%%%%%%%%%%%%%%%%%%%%%%%%%%%%%%%%%%%%%%%%%%%%%%%%%%%%%%%%%%
\paragraph{Learned Codes Visualization}
We depict in Figure \ref{fig:pc_mats} the learned codes via the visualization of the parity-check matrices. 
We can observe that for low-density codes the modifications remain small, since the code is already near local optimum, while for denser codes the change can be substantial. Also, the optimized codes tend to be more sparse than the original.

We provide in Appendix \ref{app:lso} visualizations of the \textbf{line search optimization}, demonstrating the high non-convexity and the proximity of the optimum to the current estimate. 
We provide in Appendix \ref{app:convergence} statistics on \textbf{convergence rates} and typical convergence curves demonstrating the fast and monotonic convergence.
Finally, we provide in Appendix \ref{app:bp_variant} the performance of the learned code on the efficient \textbf{Min-Sum approximation} of the BP algorithm and show that the learned code outperforms the baseline codes over the Min-Sum framework as well.

\section{Conclusions}
We present a novel gradient-based optimization method of binary linear block codes for the Belief Propagation algorithm. 
The proposed framework enables the differentiable optimization of the factor graph via weighted tensor representation. 
The optimization is efficiently carried out via a tailor-made grid search procedure that is aware of the binary constraint of the optimization problem.

A common criticism of ML-based ECC is that the neural decoder cannot be deployed directly without the application of massive deep-learning acceleration methods. Here, we show that the code can be designed efficiently in a data-driven fashion on differentiable {formulations of} classical decoders. The optimization of codes %{\color{black}, especially via its efficient initialization,} 
may open the door to the establishment of new industry standards and the creation of new families of codes.

\newpage
\bibliographystyle{iclr2025_conference}
\bibliography{references}

\begin{thebibliography}{69}
\providecommand{\natexlab}[1]{#1}
\providecommand{\url}[1]{\texttt{#1}}
\expandafter\ifx\csname urlstyle\endcsname\relax
  \providecommand{\doi}[1]{doi: #1}\else
  \providecommand{\doi}{doi: \begingroup \urlstyle{rm}\Url}\fi

\bibitem[Abu-Surra et~al.(2010)Abu-Surra, DeClercq, Divsalar, and Ryan]{abu2010trapping}
Shadi Abu-Surra, David DeClercq, Dariush Divsalar, and William~E Ryan.
\newblock Trapping set enumerators for specific ldpc codes.
\newblock In \emph{2010 Information Theory and Applications Workshop (ITA)}, pp.\  1--5. IEEE, 2010.

\bibitem[Arikan(2008)]{arikan2008channel}
Erdal Arikan.
\newblock Channel polarization: A method for constructing capacity-achieving codes.
\newblock In \emph{2008 IEEE International Symposium on Information Theory}, pp.\  1173--1177. IEEE, 2008.

\bibitem[Bengio et~al.(2013)Bengio, L{\'e}onard, and Courville]{bengio2013estimating}
Yoshua Bengio, Nicholas L{\'e}onard, and Aaron Courville.
\newblock Estimating or propagating gradients through stochastic neurons for conditional computation.
\newblock \emph{arXiv preprint arXiv:1308.3432}, 2013.

\bibitem[Bennatan et~al.(2018)Bennatan, Choukroun, and Kisilev]{bennatan2018deep}
Amir Bennatan, Yoni Choukroun, and Pavel Kisilev.
\newblock Deep learning for decoding of linear codes-a syndrome-based approach.
\newblock In \emph{2018 IEEE International Symposium on Information Theory (ISIT)}, pp.\  1595--1599. IEEE, 2018.

\bibitem[Berlekamp(1974)]{berlekamp1974key}
Elwyn~R Berlekamp.
\newblock Key papers in the development of coding theory.
\newblock \emph{(No Title)}, 1974.

\bibitem[Berrou et~al.(1993)Berrou, Glavieux, and Thitimajshima]{berrou1993near}
Claude Berrou, Alain Glavieux, and Punya Thitimajshima.
\newblock Near shannon limit error-correcting coding and decoding: Turbo-codes. 1.
\newblock In \emph{Proceedings of ICC'93-IEEE International Conference on Communications}, volume~2, pp.\  1064--1070. IEEE, 1993.

\bibitem[Boccardi et~al.(2014)Boccardi, Heath, Lozano, Marzetta, and Popovski]{boccardi2014five}
Federico Boccardi, Robert~W Heath, Angel Lozano, Thomas~L Marzetta, and Petar Popovski.
\newblock Five disruptive technology directions for 5g.
\newblock \emph{IEEE communications magazine}, 52\penalty0 (2):\penalty0 74--80, 2014.

\bibitem[Bose \& Ray-Chaudhuri(1960)Bose and Ray-Chaudhuri]{bose1960class}
Raj~Chandra Bose and Dwijendra~K Ray-Chaudhuri.
\newblock On a class of error correcting binary group codes.
\newblock \emph{Information and control}, 3\penalty0 (1):\penalty0 68--79, 1960.

\bibitem[Buchberger et~al.(2020)Buchberger, H{\"a}ger, Pfister, Schmalen, et~al.]{buchberger2020learned}
Andreas Buchberger, Christian H{\"a}ger, Henry~D Pfister, Laurent Schmalen, et~al.
\newblock Learned decimation for neural belief propagation decoders.
\newblock \emph{arXiv preprint arXiv:2011.02161}, 2020.

\bibitem[Cammerer et~al.(2017)Cammerer, Gruber, Hoydis, and ten Brink]{cammerer2017scaling}
Sebastian Cammerer, Tobias Gruber, Jakob Hoydis, and Stephan ten Brink.
\newblock Scaling deep learning-based decoding of polar codes via partitioning.
\newblock In \emph{GLOBECOM 2017-2017 IEEE Global Communications Conference}, pp.\  1--6. IEEE, 2017.

\bibitem[Choukroun \& Wolf(2022{\natexlab{a}})Choukroun and Wolf]{choukroun2022error}
Yoni Choukroun and Lior Wolf.
\newblock Error correction code transformer.
\newblock \emph{Advances in Neural Information Processing Systems (NeurIPS)}, 2022{\natexlab{a}}.

\bibitem[Choukroun \& Wolf(2022{\natexlab{b}})Choukroun and Wolf]{choukroun2022zdenoising}
Yoni Choukroun and Lior Wolf.
\newblock Denoising diffusion error correction codes.
\newblock In \emph{The Eleventh International Conference on Learning Representations}, 2022{\natexlab{b}}.

\bibitem[Choukroun \& Wolf(2024{\natexlab{a}})Choukroun and Wolf]{choukroun2024colearning}
Yoni Choukroun and Lior Wolf.
\newblock Learning linear block error correction codes.
\newblock In \emph{The Forty-first International Conference on Machine Learning}, 2024{\natexlab{a}}.

\bibitem[Choukroun \& Wolf(2024{\natexlab{b}})Choukroun and Wolf]{choukroun2024deep}
Yoni Choukroun and Lior Wolf.
\newblock Deep quantum error correction.
\newblock In \emph{Proceedings of the AAAI Conference on Artificial Intelligence}, volume~38, pp.\  64--72, 2024{\natexlab{b}}.

\bibitem[Choukroun \& Wolf(2024{\natexlab{c}})Choukroun and Wolf]{choukroun2024found}
Yoni Choukroun and Lior Wolf.
\newblock A foundation model for error correction codes.
\newblock In \emph{The Twelfth International Conference on Learning Representations}, 2024{\natexlab{c}}.
\newblock URL \url{https://openreview.net/forum?id=7KDuQPrAF3}.

\bibitem[Chow \& Liu(1968)Chow and Liu]{chow1968approximating}
CKCN Chow and Cong Liu.
\newblock Approximating discrete probability distributions with dependence trees.
\newblock \emph{IEEE transactions on Information Theory}, 14\penalty0 (3):\penalty0 462--467, 1968.

\bibitem[Chung et~al.(2001)Chung, Forney, Richardson, and Urbanke]{chung2001design}
Sae-Young Chung, G~David Forney, Thomas~J Richardson, and R{\"u}diger Urbanke.
\newblock On the design of low-density parity-check codes within 0.0045 db of the shannon limit.
\newblock \emph{IEEE Communications letters}, 5\penalty0 (2):\penalty0 58--60, 2001.

\bibitem[Costello \& Forney(2007)Costello and Forney]{costello2007channel}
Daniel~J Costello and G~David Forney.
\newblock Channel coding: The road to channel capacity.
\newblock \emph{Proceedings of the IEEE}, 95\penalty0 (6):\penalty0 1150--1177, 2007.

\bibitem[Courbariaux et~al.(2016)Courbariaux, Hubara, Soudry, El-Yaniv, and Bengio]{courbariaux2016binarized}
Matthieu Courbariaux, Itay Hubara, Daniel Soudry, Ran El-Yaniv, and Yoshua Bengio.
\newblock Binarized neural networks: Training deep neural networks with weights and activations constrained to+ 1 or-1.
\newblock \emph{arXiv preprint arXiv:1602.02830}, 2016.

\bibitem[De~Cola et~al.(2011)De~Cola, Paolini, Liva, and Calzolari]{de2011reliability}
Tomaso De~Cola, Enrico Paolini, Gianluigi Liva, and Gian~Paolo Calzolari.
\newblock Reliability options for data communications in the future deep-space missions.
\newblock \emph{Proceedings of the IEEE}, 99\penalty0 (11):\penalty0 2056--2074, 2011.

\bibitem[Durisi et~al.(2016)Durisi, Koch, and Popovski]{durisi2016toward}
Giuseppe Durisi, Tobias Koch, and Petar Popovski.
\newblock Toward massive, ultrareliable, and low-latency wireless communication with short packets.
\newblock \emph{Proceedings of the IEEE}, 104\penalty0 (9):\penalty0 1711--1726, 2016.

\bibitem[Eleftheriou \& Olcer(2002)Eleftheriou and Olcer]{eleftheriou2002low}
Evangelos Eleftheriou and Sedat Olcer.
\newblock Low-density parity-check codes for digital subscriber lines.
\newblock In \emph{2002 IEEE International Conference on Communications. Conference Proceedings. ICC 2002 (Cat. No. 02CH37333)}, volume~3, pp.\  1752--1757. IEEE, 2002.

\bibitem[Elkelesh et~al.(2019)Elkelesh, Ebada, Cammerer, Schmalen, and Ten~Brink]{elkelesh2019decoder}
Ahmed Elkelesh, Moustafa Ebada, Sebastian Cammerer, Laurent Schmalen, and Stephan Ten~Brink.
\newblock Decoder-in-the-loop: Genetic optimization-based ldpc code design.
\newblock \emph{IEEE access}, 7:\penalty0 141161--141170, 2019.

\bibitem[ESTI(2021)]{ETSI}
ESTI.
\newblock 5g nr multiplexing and channel coding. etsi 3gpp ts 38.212.
\newblock \url{https://www.etsi.org/deliver/etsi_ts/138200_138299/138212/16.02.00_60/ts_138212v160200p.pdf}, 2021.

\bibitem[Forney(1966)]{forney1966concatenated}
GD~Forney.
\newblock Concatenated codes. cambridge.
\newblock \emph{Massachusetts: Massachusetts Institute of Technology}, 1966.

\bibitem[Gallager(1962)]{gallager1962low}
Robert Gallager.
\newblock Low-density parity-check codes.
\newblock \emph{IRE Transactions on information theory}, 8\penalty0 (1):\penalty0 21--28, 1962.

\bibitem[Gruber et~al.(2017)Gruber, Cammerer, Hoydis, and ten Brink]{gruber2017deep}
Tobias Gruber, Sebastian Cammerer, Jakob Hoydis, and Stephan ten Brink.
\newblock On deep learning-based channel decoding.
\newblock In \emph{2017 51st Annual Conference on Information Sciences and Systems (CISS)}, pp.\  1--6. IEEE, 2017.

\bibitem[Helmling et~al.(2019)Helmling, Scholl, Gensheimer, Dietz, Kraft, Ruzika, and Wehn]{channelcodes}
Michael Helmling, Stefan Scholl, Florian Gensheimer, Tobias Dietz, Kira Kraft, Stefan Ruzika, and Norbert Wehn.
\newblock {D}atabase of {C}hannel {C}odes and {ML} {S}imulation {R}esults.
\newblock \url{www.uni-kl.de/channel-codes}, 2019.

\bibitem[Hoydis et~al.(2022)Hoydis, Cammerer, {Ait Aoudia}, Vem, Binder, Marcus, and Keller]{sionna}
Jakob Hoydis, Sebastian Cammerer, Fayçal {Ait Aoudia}, Avinash Vem, Nikolaus Binder, Guillermo Marcus, and Alexander Keller.
\newblock Sionna: An open-source library for next-generation physical layer research.
\newblock \emph{arXiv preprint}, Mar. 2022.

\bibitem[Hu et~al.(2001)Hu, Eleftheriou, and Arnold]{hu2001progressive}
Xiao-Yu Hu, Evangelos Eleftheriou, and D-M Arnold.
\newblock Progressive edge-growth tanner graphs.
\newblock In \emph{GLOBECOM'01. IEEE Global Telecommunications Conference (Cat. No. 01CH37270)}, volume~2, pp.\  995--1001. IEEE, 2001.

\bibitem[Hu et~al.(2005)Hu, Eleftheriou, and Arnold]{hu2005regular}
Xiao-Yu Hu, Evangelos Eleftheriou, and Dieter-Michael Arnold.
\newblock Regular and irregular progressive edge-growth tanner graphs.
\newblock \emph{IEEE transactions on information theory}, 51\penalty0 (1):\penalty0 386--398, 2005.

\bibitem[Jiang et~al.(2019{\natexlab{a}})Jiang, Kannan, Kim, Oh, Asnani, and Viswanath]{jiang2019deepturbo}
Yihan Jiang, Sreeram Kannan, Hyeji Kim, Sewoong Oh, Himanshu Asnani, and Pramod Viswanath.
\newblock Deepturbo: Deep turbo decoder.
\newblock In \emph{2019 IEEE 20th International Workshop on Signal Processing Advances in Wireless Communications (SPAWC)}, pp.\  1--5. IEEE, 2019{\natexlab{a}}.

\bibitem[Jiang et~al.(2019{\natexlab{b}})Jiang, Kim, Asnani, Kannan, Oh, and Viswanath]{jiang2019turbo}
Yihan Jiang, Hyeji Kim, Himanshu Asnani, Sreeram Kannan, Sewoong Oh, and Pramod Viswanath.
\newblock Turbo autoencoder: Deep learning based channel codes for point-to-point communication channels.
\newblock \emph{Advances in neural information processing systems}, 32, 2019{\natexlab{b}}.

\bibitem[Jin et~al.(2000)Jin, Khandekar, McEliece, et~al.]{jin2000irregular}
Hui Jin, Aamod Khandekar, Robert McEliece, et~al.
\newblock Irregular repeat-accumulate codes.
\newblock In \emph{Proc. 2nd Int. Symp. Turbo codes and related topics}, pp.\  1--8. Citeseer, 2000.

\bibitem[Kim et~al.(2018{\natexlab{a}})Kim, Jiang, Kannan, Oh, and Viswanath]{kim2018deepcode}
Hyeji Kim, Yihan Jiang, Sreeram Kannan, Sewoong Oh, and Pramod Viswanath.
\newblock Deepcode: Feedback codes via deep learning.
\newblock In \emph{Advances in Neural Information Processing Systems (NIPS)}, pp.\  9436--9446, 2018{\natexlab{a}}.

\bibitem[Kim et~al.(2018{\natexlab{b}})Kim, Jiang, Rana, Kannan, Oh, and Viswanath]{kim2018communication}
Hyeji Kim, Yihan Jiang, Ranvir Rana, Sreeram Kannan, Sewoong Oh, and Pramod Viswanath.
\newblock Communication algorithms via deep learning.
\newblock In \emph{Sixth International Conference on Learning Representations (ICLR)}, 2018{\natexlab{b}}.

\bibitem[Koller \& Friedman(2009)Koller and Friedman]{koller2009probabilistic}
Daphne Koller and Nir Friedman.
\newblock \emph{Probabilistic graphical models: principles and techniques}.
\newblock MIT press, 2009.

\bibitem[Kou et~al.(2001)Kou, Lin, and Fossorier]{kou2001low}
Yu~Kou, Shu Lin, and Marc~PC Fossorier.
\newblock Low-density parity-check codes based on finite geometries: a rediscovery and new results.
\newblock \emph{IEEE Transactions on Information theory}, 47\penalty0 (7):\penalty0 2711--2736, 2001.

\bibitem[Kudekar et~al.(2011)Kudekar, Richardson, and Urbanke]{kudekar2011threshold}
Shrinivas Kudekar, Thomas~J Richardson, and R{\"u}diger~L Urbanke.
\newblock Threshold saturation via spatial coupling: Why convolutional ldpc ensembles perform so well over the bec.
\newblock \emph{IEEE Transactions on Information Theory}, 57\penalty0 (2):\penalty0 803--834, 2011.

\bibitem[Kwak et~al.(2023)Kwak, Yun, Kim, Kim, and No]{kwak2023boosting}
Hee-Youl Kwak, Dae-Young Yun, Yongjune Kim, Sang-Hyo Kim, and Jong-Seon No.
\newblock Boosting learning for ldpc codes to improve the error-floor performance.
\newblock \emph{arXiv preprint arXiv:2310.07194}, 2023.

\bibitem[Liva et~al.(2016)Liva, Gaudio, Ninacs, and Jerkovits]{liva2016code}
Gianluigi Liva, Lorenzo Gaudio, Tudor Ninacs, and Thomas Jerkovits.
\newblock Code design for short blocks: A survey.
\newblock \emph{arXiv preprint arXiv:1610.00873}, 2016.

\bibitem[Luby et~al.(2001)Luby, Mitzenmacher, Shokrollahi, and Spielman]{luby2001improved}
Michael~G Luby, Michael Mitzenmacher, Mohammad~Amin Shokrollahi, and Daniel~A Spielman.
\newblock Improved low-density parity-check codes using irregular graphs.
\newblock \emph{IEEE Transactions on information Theory}, 47\penalty0 (2):\penalty0 585--598, 2001.

\bibitem[Lucas et~al.(2000)Lucas, Fossorier, Kou, and Lin]{lucas2000iterative}
Rainer Lucas, Marc~PC Fossorier, Yu~Kou, and Shu Lin.
\newblock Iterative decoding of one-step majority logic deductible codes based on belief propagation.
\newblock \emph{IEEE Transactions on Communications}, 48\penalty0 (6):\penalty0 931--937, 2000.

\bibitem[Lugosch \& Gross(2017)Lugosch and Gross]{lugosch2017neural}
Loren Lugosch and Warren~J Gross.
\newblock Neural offset min-sum decoding.
\newblock In \emph{2017 IEEE International Symposium on Information Theory (ISIT)}, pp.\  1361--1365. IEEE, 2017.

\bibitem[MacKay()]{pge_imp}
David MacKay.
\newblock Progressive edge growth implementation.
\newblock \url{https://inference.org.uk/mackay/PEG_ECC.html}.

\bibitem[MacKay(1999)]{mackay1999good}
David~JC MacKay.
\newblock Good error-correcting codes based on very sparse matrices.
\newblock \emph{IEEE transactions on Information Theory}, 45\penalty0 (2):\penalty0 399--431, 1999.

\bibitem[MacKay \& Neal(1995)MacKay and Neal]{mackay1995good}
David~JC MacKay and Radford~M Neal.
\newblock Good codes based on very sparse matrices.
\newblock In \emph{IMA International Conference on Cryptography and Coding}, pp.\  100--111. Springer, 1995.

\bibitem[Nachmani \& Wolf(2019)Nachmani and Wolf]{nachmani2019hyper}
Eliya Nachmani and Lior Wolf.
\newblock Hyper-graph-network decoders for block codes.
\newblock In \emph{Advances in Neural Information Processing Systems}, pp.\  2326--2336, 2019.

\bibitem[Nachmani \& Wolf(2021)Nachmani and Wolf]{nachmani2021autoregressive}
Eliya Nachmani and Lior Wolf.
\newblock Autoregressive belief propagation for decoding block codes.
\newblock \emph{arXiv preprint arXiv:2103.11780}, 2021.

\bibitem[Nachmani et~al.(2016)Nachmani, Be'ery, and Burshtein]{nachmani2016learning}
Eliya Nachmani, Yair Be'ery, and David Burshtein.
\newblock Learning to decode linear codes using deep learning.
\newblock In \emph{2016 54th Annual Allerton Conference on Communication, Control, and Computing (Allerton)}, pp.\  341--346. IEEE, 2016.

\bibitem[Nachmani et~al.(2018)Nachmani, Marciano, Lugosch, Gross, Burshtein, and Be’ery]{nachmani2017learning}
Eliya Nachmani, Elad Marciano, Loren Lugosch, Warren~J Gross, David Burshtein, and Yair Be’ery.
\newblock Deep learning methods for improved decoding of linear codes.
\newblock \emph{IEEE Journal of Selected Topics in Signal Processing}, 12\penalty0 (1):\penalty0 119--131, 2018.

\bibitem[Narayanaswami(2001)]{narayanaswami2001coded}
Ravi Narayanaswami.
\newblock \emph{Coded modulation with low density parity check codes}.
\newblock PhD thesis, Texas A\&M University, 2001.

\bibitem[Nocedal \& Wright(2006)Nocedal and Wright]{Nocedal2006}
Jorge Nocedal and Stephen~J. Wright.
\newblock \emph{Line Search Methods}, pp.\  30--65.
\newblock Springer New York, New York, NY, 2006.
\newblock ISBN 978-0-387-40065-5.
\newblock \doi{10.1007/978-0-387-40065-5_3}.
\newblock URL \url{https://doi.org/10.1007/978-0-387-40065-5_3}.

\bibitem[O'Shea \& Hoydis(2017)O'Shea and Hoydis]{AutoencoderComm}
Timothy~J O'Shea and Jakob Hoydis.
\newblock An introduction to machine learning communications systems.
\newblock \emph{arXiv preprint arXiv:1702.00832}, 2017.

\bibitem[Paolini et~al.(2015)Paolini, Stefanovic, Liva, and Popovski]{paolini2015coded}
Enrico Paolini, Cedomir Stefanovic, Gianluigi Liva, and Petar Popovski.
\newblock Coded random access: Applying codes on graphs to design random access protocols.
\newblock \emph{IEEE Communications Magazine}, 53\penalty0 (6):\penalty0 144--150, 2015.

\bibitem[Pearl(1988)]{pearl1988probabilistic}
Judea Pearl.
\newblock \emph{Probabilistic reasoning in intelligent systems: networks of plausible inference}.
\newblock Morgan kaufmann, 1988.

\bibitem[Rastegari et~al.(2016)Rastegari, Ordonez, Redmon, and Farhadi]{xnor_net}
Mohammad Rastegari, Vicente Ordonez, Joseph Redmon, and Ali Farhadi.
\newblock Xnor-net: Imagenet classification using binary convolutional neural networks.
\newblock In \emph{European Conference on Computer Vision}, pp.\  525--542. Springer, 2016.

\bibitem[Raviv et~al.(2020)Raviv, Caciularu, Raviv, Goldberger, and Be'ery]{raviv2020graph}
Nir Raviv, Avi Caciularu, Tomer Raviv, Jacob Goldberger, and Yair Be'ery.
\newblock perm2vec: Graph permutation selection for decoding of error correction codes using self-attention.
\newblock \emph{arXiv preprint arXiv:2002.02315}, 2020.

\bibitem[Raviv et~al.(2023)Raviv, Goldmann, Vayner, Be'ery, and Shlezinger]{raviv2023crc}
Tomer Raviv, Alon Goldmann, Ofek Vayner, Yair Be'ery, and Nir Shlezinger.
\newblock Crc-aided learned ensembles of belief-propagation polar decoders.
\newblock \emph{arXiv preprint arXiv:2301.06060}, 2023.

\bibitem[Reed \& Solomon(1960)Reed and Solomon]{reed1960polynomial}
Irving~S Reed and Gustave Solomon.
\newblock Polynomial codes over certain finite fields.
\newblock \emph{Journal of the society for industrial and applied mathematics}, 8\penalty0 (2):\penalty0 300--304, 1960.

\bibitem[Richardson \& Urbanke(2001)Richardson and Urbanke]{richardson2001capacity}
Thomas~J Richardson and R{\"u}diger~L Urbanke.
\newblock The capacity of low-density parity-check codes under message-passing decoding.
\newblock \emph{IEEE Transactions on information theory}, 47\penalty0 (2):\penalty0 599--618, 2001.

\bibitem[Richardson et~al.(2001)Richardson, Shokrollahi, and Urbanke]{richardson2001design}
Thomas~J Richardson, Mohammad~Amin Shokrollahi, and R{\"u}diger~L Urbanke.
\newblock Design of capacity-approaching irregular low-density parity-check codes.
\newblock \emph{IEEE transactions on information theory}, 47\penalty0 (2):\penalty0 619--637, 2001.

\bibitem[Shannon(1948)]{shannon1948mathematical}
Claude~Elwood Shannon.
\newblock A mathematical theory of communication.
\newblock \emph{The Bell system technical journal}, 27\penalty0 (3):\penalty0 379--423, 1948.

\bibitem[Tanner(1981)]{tanner1981recursive}
R~Tanner.
\newblock A recursive approach to low complexity codes.
\newblock \emph{IEEE Transactions on information theory}, 27\penalty0 (5):\penalty0 533--547, 1981.

\bibitem[Tian(2013)]{tian2013branch}
Jin Tian.
\newblock A branch-and-bound algorithm for mdl learning bayesian networks.
\newblock \emph{arXiv preprint arXiv:1301.3897}, 2013.

\bibitem[Vasic \& Milenkovic(2004)Vasic and Milenkovic]{vasic2004combinatorial}
Bane Vasic and Olgica Milenkovic.
\newblock Combinatorial constructions of low-density parity-check codes for iterative decoding.
\newblock \emph{IEEE Transactions on information theory}, 50\penalty0 (6):\penalty0 1156--1176, 2004.

\bibitem[Wolf(1978)]{wolf1978efficient}
Jack Wolf.
\newblock Efficient maximum likelihood decoding of linear block codes using a trellis.
\newblock \emph{IEEE Transactions on Information Theory}, 24\penalty0 (1):\penalty0 76--80, 1978.

\bibitem[Yang et~al.(2004)Yang, Ryan, and Li]{yang2004design}
Michael Yang, William~E Ryan, and Yan Li.
\newblock Design of efficiently encodable moderate-length high-rate irregular ldpc codes.
\newblock \emph{IEEE Transactions on Communications}, 52\penalty0 (4):\penalty0 564--571, 2004.

\bibitem[Yin et~al.(2019)Yin, Lyu, Zhang, Osher, Qi, and Xin]{yin2019understanding}
Penghang Yin, Jiancheng Lyu, Shuai Zhang, Stanley Osher, Yingyong Qi, and Jack Xin.
\newblock Understanding straight-through estimator in training activation quantized neural nets.
\newblock \emph{arXiv preprint arXiv:1903.05662}, 2019.

\end{thebibliography}

%%%%%%%%%%%%%%%%%%%%%%%%%%%%%%%%%%%%%%%%%%%%%%%%%%%%%%%%%%%%
\newpage
\clearpage
\appendix

\section{Hyper-Parameter Tuning}
\label{app:hyperparam}
{\color{black}
Under the problem's stochastic optimization, we provide here the different modifications used to obtain better performance.
The first set of training/optimization hyperparameters is the $E_{b}/N_{0}$ range defined as $(u,7)$ with $u\in \{3,4,5\}$.
The second set of hyperparameters is the data sampling, where we experimented with random data (i.e., classical setting) and data with non-zero syndromes only.
Finally, for better backpropagation, we also experimented with a soft approximation $\tilde{H}$ of the binary $H$ during the optimization, defined as
  \[
    \tilde{H}_{ij}=\left\{
                \begin{array}{ll}
                  (-1)^{z}\epsilon, \ \ \ \ \ \ \ if \ \ \ H_{ij}=0 \\
                  1, \ \ \ \ \ \ \ else
                \end{array}
              \right.
  \]
  where $z\sim Bern(0.5)$ and $\epsilon$ is a small scalar ($10^{-7}$ in our experiments).
  We note we only used a size 15 random subset of all the possible permutations of the hyperparameters mentioned above.
}

\section{More SNR results}
\label{app:more_snr}
We provide results on a larger range of SNRs in Table \ref{tab:main_ber}.
\begin{table}[h]
    \centering
    \caption{
    A comparison of the negative natural logarithm of Bit Error Rate (BER) for several normalized SNR values of our method with classical codes. Higher is better. BP results are provided for 5 iterations in the first row and 15 in the second row. The best results are in bold.}
    \label{tab:main_ber}
    % \setlength\tabcolsep{1.5pt}
%    \smallskip
% \vspace{-2mm}
    \resizebox{0.99985\textwidth}{!}{%
    \begin{tabular}{@{}lc@{~}c@{~}c@{~}c@{~}cc@{~}c@{~}c@{~}c@{~}cc@{~}c@{~}c@{~}c@{~}cc@{~}c@{~}c@{~}c@{~}cc@{~}c@{~}c@{~}c@{~}cc@{~}c@{~}c@{~}c@{~}c@{}}
    \toprule
        Channel & \multicolumn{10}{c}{AWGN} & \multicolumn{10}{c}{Fading} & \multicolumn{10}{c}{Bursting}\\
  %       \cmidrule(lr){2-4}
  %       \cmidrule(lr){5-13}
		% \cmidrule(lr){14-16}
        Method & \multicolumn{5}{c}{BP} & \multicolumn{5}{c}{Our} & \multicolumn{5}{c}{BP} & \multicolumn{5}{c}{Our} &  \multicolumn{5}{c}{BP} & \multicolumn{5}{c}{Our} \\
  %       \cmidrule(lr){2-4}
  %       \cmidrule(lr){5-7}
  %       \cmidrule(lr){8-10}
  %       \cmidrule(lr){11-13}
		% \cmidrule(lr){14-16}
         $E_{b}/N_{0}$ & 3 & 4 & 5 & 6 & 7 & 3 & 4 & 5 & 6 & 7 & 3 & 4 & 5 & 6 & 7 & 3 & 4 & 5 & 6 & 7 & 3 & 4 & 5 & 6 & 7 & 3 & 4 & 5 & 6 & 7  \\ 
         \midrule                                                                                                                                                                                                                                                                                                                                                                                                                                      
		%%%%%%%%%%%%%%%%%%%
  BCH(63,45)      	\comment{AWGN} \comment{BP} & \makecell{3.35\\3.40} & \makecell{4.06\\4.21} & \makecell{4.91\\5.24} & \makecell{6.04\\6.59} & \makecell{7.47\\8.35} 				\comment{Our} & \makecell{\bf4.23\\ \bf4.36} & \makecell{\bf5.44\\ \bf5.70} & \makecell{\bf6.93\\ \bf7.35} & \makecell{\bf8.60\\ \bf9.16} & \makecell{\bf10.27\\ \bf11.10} 					\comment{Fading_1.0} \comment{BP} & \makecell{2.77\\2.79} & \makecell{3.09\\3.13} & \makecell{3.46\\3.55} & \makecell{3.90\\4.04} & \makecell{4.37\\4.61} 				\comment{Our} & \makecell{\bf3.42\\ \bf3.50} & \makecell{\bf3.96\\ \bf4.10} & \makecell{\bf4.58\\ \bf4.80} & \makecell{\bf5.27\\ \bf5.56} & \makecell{\bf5.99\\ \bf6.36} 						\comment{Bursting} \comment{BP} & \makecell{3.00\\3.02} & \makecell{3.60\\3.67} & \makecell{4.32\\4.52} & \makecell{5.19\\5.59} & \makecell{6.25\\6.93} 					\comment{Our} & \makecell{\bf3.24\\ \bf3.31} & \makecell{\bf4.05\\ \bf4.21} & \makecell{\bf5.07\\ \bf5.40} & \makecell{\bf6.27\\ \bf6.85} & \makecell{\bf7.67\\ \bf8.55} \\ \midrule

CCSDS(128,64)   	\comment{AWGN} \comment{BP} & \makecell{4.32\\4.82} & \makecell{6.46\\7.32} & \makecell{9.61\\10.83} & \makecell{13.99\\15.43} & \makecell{\bf18.27\\ \bf18.51} 	\comment{Our} & \makecell{\bf4.99\\ \bf5.80} & \makecell{\bf7.34\\ \bf8.61} & \makecell{\bf10.48\\ \bf12.26} & \makecell{\bf14.37\\ \bf16.00} & \makecell{17.38\\18.15} 					\comment{Fading_1.0} \comment{BP} & \makecell{4.37\\4.89} & \makecell{5.72\\6.43} & \makecell{7.42\\8.29} & \makecell{9.47\\10.28} & \makecell{11.84\\12.88} 			\comment{Our} & \makecell{\bf5.22\\ \bf6.22} & \makecell{\bf6.73\\ \bf8.05} & \makecell{\bf8.45\\ \bf10.07} & \makecell{\bf10.45\\ \bf12.37} & \makecell{\bf12.36\\ \bf14.87} 					\comment{Bursting} \comment{BP} & \makecell{3.62\\3.97} & \makecell{5.29\\5.98} & \makecell{7.81\\8.85} & \makecell{11.25\\12.53} & \makecell{\bf15.59\\ \bf17.10} 			\comment{Our} & \makecell{\bf4.32\\ \bf5.05} & \makecell{\bf6.23\\ \bf7.39} & \makecell{\bf8.80\\ \bf10.43} & \makecell{\bf11.90\\ \bf13.28} & \makecell{14.76\\15.56} \\ \midrule

LDPC(121,60)    	\comment{AWGN} \comment{BP} & \makecell{3.33\\3.53} & \makecell{4.81\\5.31} & \makecell{7.17\\7.96} & \makecell{10.75\\11.85} & \makecell{15.69\\17.01} 			\comment{Our} & \makecell{\bf5.29\\ \bf6.18} & \makecell{\bf7.70\\ \bf8.86} & \makecell{\bf10.87\\ \bf11.91} & \makecell{\bf14.25\\ \bf14.41} & \makecell{\bf16.82\\ \bf17.04} 				\comment{Fading_1.0} \comment{BP} & \makecell{3.24\\3.44} & \makecell{4.10\\4.42} & \makecell{5.23\\5.61} & \makecell{6.68\\7.04} & \makecell{8.56\\8.77} 				\comment{Our} & \makecell{\bf5.18\\ \bf6.04} & \makecell{\bf6.68\\ \bf7.71} & \makecell{\bf8.47\\ \bf9.67} & \makecell{\bf10.50\\ \bf11.76} & \makecell{\bf12.31\\ \bf14.21} 					\comment{Bursting} \comment{BP} & \makecell{2.87\\2.98} & \makecell{3.97\\4.31} & \makecell{5.75\\6.37} & \makecell{8.40\\9.25} & \makecell{12.16\\13.10} 					\comment{Our} & \makecell{\bf4.32\\ \bf5.02} & \makecell{\bf6.23\\ \bf7.26} & \makecell{\bf8.89\\ \bf10.03} & \makecell{\bf11.98\\ \bf12.88} & \makecell{\bf14.91\\ \bf15.14} \\ \midrule

LDPC(121,80)    	\comment{AWGN} \comment{BP} & \makecell{4.50\\4.85} & \makecell{6.59\\7.35} & \makecell{9.68\\10.94} & \makecell{13.43\\15.46} & \makecell{\bf18.51\\ \bf19.61} 	\comment{Our} & \makecell{\bf5.28\\ \bf5.85} & \makecell{\bf7.77\\ \bf8.75} & \makecell{\bf11.21\\ \bf12.45} & \makecell{\bf15.06\\ \bf15.67} & \makecell{18.40\\18.30} 					\comment{Fading_1.0} \comment{BP} & \makecell{3.67\\3.89} & \makecell{4.60\\4.97} & \makecell{5.80\\6.29} & \makecell{7.22\\7.82} & \makecell{8.95\\9.58} 				\comment{Our} & \makecell{\bf4.42\\ \bf4.89} & \makecell{\bf5.55\\ \bf6.25} & \makecell{\bf6.90\\ \bf7.80} & \makecell{\bf8.36\\ \bf9.47} & \makecell{\bf10.00\\ \bf11.21} 						\comment{Bursting} \comment{BP} & \makecell{3.74\\3.94} & \makecell{5.30\\5.81} & \makecell{7.60\\8.50} & \makecell{10.66\\12.15} & \makecell{14.88\\16.45} 				\comment{Our} & \makecell{\bf4.32\\ \bf4.73} & \makecell{\bf6.23\\ \bf6.99} & \makecell{\bf8.87\\ \bf10.09} & \makecell{\bf12.19\\ \bf13.74} & \makecell{\bf16.31\\ \bf17.55} \\ \midrule

LDPC(128,64)    	\comment{AWGN} \comment{BP} & \makecell{2.88\\3.04} & \makecell{3.66\\4.00} & \makecell{4.65\\5.16} & \makecell{5.80\\6.42} & \makecell{7.03\\7.77} 				\comment{Our} & \makecell{\bf4.07\\ \bf4.71} & \makecell{\bf5.54\\ \bf6.56} & \makecell{\bf7.37\\ \bf8.70} & \makecell{\bf9.44\\ \bf10.81} & \makecell{\bf11.71\\ \bf12.92} 				\comment{Fading_1.0} \comment{BP} & \makecell{2.73\\2.90} & \makecell{3.22\\3.51} & \makecell{3.80\\4.18} & \makecell{4.44\\4.84} & \makecell{5.14\\5.54} 				\comment{Our} & \makecell{\bf3.95\\ \bf4.55} & \makecell{\bf4.86\\ \bf5.64} & \makecell{\bf5.94\\ \bf6.85} & \makecell{\bf7.15\\ \bf8.14} & \makecell{\bf8.38\\ \bf9.59} 						\comment{Bursting} \comment{BP} & \makecell{2.58\\2.68} & \makecell{3.23\\3.48} & \makecell{4.08\\4.51} & \makecell{5.09\\5.66} & \makecell{6.21\\6.88} 					\comment{Our} & \makecell{\bf2.80\\ \bf2.97} & \makecell{\bf3.72\\ \bf4.13} & \makecell{\bf5.00\\ \bf5.72} & \makecell{\bf6.54\\ \bf7.66} & \makecell{\bf8.30\\ \bf9.86} \\ \midrule

LDPC(32,16)     	\comment{AWGN} \comment{BP} & \makecell{3.45\\3.59} & \makecell{4.36\\4.64} & \makecell{5.59\\6.07} & \makecell{7.18\\7.94} & \makecell{9.19\\10.23} 				\comment{Our} & \makecell{\bf4.29\\ \bf4.47} & \makecell{\bf5.48\\ \bf5.76} & \makecell{\bf7.02\\ \bf7.44} & \makecell{\bf8.92\\ \bf9.41} & \makecell{\bf11.23\\ \bf12.03} 					\comment{Fading_1.0} \comment{BP} & \makecell{3.44\\3.62} & \makecell{4.03\\4.29} & \makecell{4.70\\5.06} & \makecell{5.47\\5.90} & \makecell{6.31\\6.83} 				\comment{Our} & \makecell{\bf4.53\\ \bf4.67} & \makecell{\bf5.26\\ \bf5.43} & \makecell{\bf6.02\\ \bf6.23} & \makecell{\bf6.82\\ \bf6.97} & \makecell{\bf7.61\\ \bf7.81} 						\comment{Bursting} \comment{BP} & \makecell{3.10\\3.21} & \makecell{3.88\\4.09} & \makecell{4.89\\5.26} & \makecell{6.18\\6.76} & \makecell{7.82\\8.58} 					\comment{Our} & \makecell{\bf3.78\\ \bf3.93} & \makecell{\bf4.77\\ \bf5.01} & \makecell{\bf6.02\\ \bf6.35} & \makecell{\bf7.52\\ \bf7.96} & \makecell{\bf9.22\\ \bf9.72} \\ \midrule

LDPC(96,48)     	\comment{AWGN} \comment{BP} & \makecell{4.72\\5.20} & \makecell{6.73\\7.50} & \makecell{9.48\\10.61} & \makecell{12.98\\ \bf14.26} & \makecell{\bf16.87\\ \bf17.80} \comment{Our} & \makecell{\bf5.17\\ \bf5.85} & \makecell{\bf7.22\\ \bf8.29} & \makecell{\bf9.96\\ \bf11.12} & \makecell{\bf13.37\\14.06} & \makecell{16.45\\17.19} 							\comment{Fading_1.0} \comment{BP} & \makecell{3.19\\3.44} & \makecell{3.83\\4.17} & \makecell{4.57\\4.94} & \makecell{5.35\\5.73} & \makecell{6.17\\6.58} 				\comment{Our} & \makecell{\bf4.38\\ \bf4.99} & \makecell{\bf5.37\\ \bf6.14} & \makecell{\bf6.51\\ \bf7.38} & \makecell{\bf7.71\\ \bf8.65} & \makecell{\bf8.94\\ \bf9.77} 						\comment{Bursting} \comment{BP} & \makecell{4.03\\4.40} & \makecell{5.68\\6.33} & \makecell{7.94\\8.91} & \makecell{10.90\\ \bf11.99} & \makecell{\bf14.27\\ \bf15.55} 		\comment{Our} & \makecell{\bf4.23\\ \bf4.71} & \makecell{\bf5.90\\ \bf6.71} & \makecell{\bf8.19\\ \bf9.28} & \makecell{\bf10.91\\11.75} & \makecell{13.61\\14.06} \\ \midrule

LTE(132,40)     	\comment{AWGN} \comment{BP} & \makecell{2.49\\2.85} & \makecell{2.94\\3.37} & \makecell{3.32\\3.79} & \makecell{3.57\\4.09} & \makecell{3.81\\4.32} 				\comment{Our} & \makecell{\bf2.72\\ \bf3.26} & \makecell{\bf3.25\\ \bf3.93} & \makecell{\bf3.71\\ \bf4.49} & \makecell{\bf4.04\\ \bf4.89} & \makecell{\bf4.36\\ \bf5.22} 					\comment{Fading_1.0} \comment{BP} & \makecell{2.82\\3.29} & \makecell{3.17\\3.60} & \makecell{3.45\\3.82} & \makecell{3.67\\4.01} & \makecell{3.89\\4.21} 				\comment{Our} & \makecell{\bf3.97\\ \bf4.78} & \makecell{\bf4.49\\ \bf5.32} & \makecell{\bf4.99\\ \bf5.81} & \makecell{\bf5.47\\ \bf6.31} & \makecell{\bf5.96\\ \bf6.84} 						\comment{Bursting} \comment{BP} & \makecell{2.30\\2.63} & \makecell{2.75\\3.17} & \makecell{3.17\\3.62} & \makecell{3.47\\3.96} & \makecell{3.70\\4.19} 					\comment{Our} & \makecell{\bf2.48\\ \bf2.92} & \makecell{\bf2.99\\ \bf3.53} & \makecell{\bf3.44\\ \bf4.03} & \makecell{\bf3.78\\ \bf4.41} & \makecell{\bf4.05\\ \bf4.70} \\ \midrule

MACKAY(96,48)   	\comment{AWGN} \comment{BP} & \makecell{4.77\\5.28} & \makecell{6.75\\7.59} & \makecell{9.45\\10.52} & \makecell{\bf12.85\\ \bf14.09} & \makecell{\bf16.37\\17.43} \comment{Our} & \makecell{\bf5.03\\ \bf5.63} & \makecell{\bf7.03\\ \bf7.99} & \makecell{\bf9.63\\ \bf10.97} & \makecell{12.78\\14.05} & \makecell{16.11\\ \bf17.49} 							\comment{Fading_1.0} \comment{BP} & \makecell{4.98\\5.55} & \makecell{6.28\\7.04} & \makecell{7.86\\8.76} & \makecell{9.55\\10.64} & \makecell{11.30\\12.58} 			\comment{Our} & \makecell{\bf5.18\\ \bf5.93} & \makecell{\bf6.53\\ \bf7.47} & \makecell{\bf8.06\\ \bf9.32} & \makecell{\bf9.77\\ \bf11.19} & \makecell{\bf11.58\\ \bf13.14} 					\comment{Bursting} \comment{BP} & \makecell{4.08\\4.47} & \makecell{5.72\\6.39} & \makecell{7.97\\8.90} & \makecell{10.81\\11.91} & \makecell{13.92\\15.23} 				\comment{Our} & \makecell{\bf4.28\\ \bf4.79} & \makecell{\bf5.95\\ \bf6.82} & \makecell{\bf8.23\\ \bf9.41} & \makecell{\bf10.91\\ \bf12.71} & \makecell{\bf14.02\\ \bf15.81} \\ \midrule

POLAR(128,86)   	\comment{AWGN} \comment{BP} & \makecell{3.25\\3.36} & \makecell{3.76\\4.02} & \makecell{4.17\\4.67} & \makecell{4.58\\5.38} & \makecell{5.12\\6.19} 				\comment{Our} & \makecell{\bf3.72\\ \bf3.96} & \makecell{\bf4.83\\ \bf5.37} & \makecell{\bf5.87\\ \bf6.88} & \makecell{\bf6.58\\ \bf8.10} & \makecell{\bf7.21\\ \bf9.00} 					\comment{Fading_1.0} \comment{BP} & \makecell{2.80\\2.87} & \makecell{3.15\\3.28} & \makecell{3.53\\3.73} & \makecell{3.91\\4.18} & \makecell{4.26\\4.60} 				\comment{Our} & \makecell{\bf3.10\\ \bf3.26} & \makecell{\bf3.64\\ \bf3.92} & \makecell{\bf4.28\\ \bf4.70} & \makecell{\bf4.94\\ \bf5.52} & \makecell{\bf5.58\\ \bf6.30} 						\comment{Bursting} \comment{BP} & \makecell{2.95\\3.02} & \makecell{3.48\\3.65} & \makecell{3.96\\4.31} & \makecell{4.37\\4.97} & \makecell{4.78\\5.66} 					\comment{Our} & \makecell{\bf2.96\\ \bf3.03} & \makecell{\bf3.69\\ \bf3.87} & \makecell{\bf4.51\\ \bf4.91} & \makecell{\bf5.18\\ \bf5.91} & \makecell{\bf5.73\\ \bf6.88} \\ \midrule

RS(60,52)       	\comment{AWGN} \comment{BP} & \makecell{3.65\\3.70} & \makecell{4.41\\4.54} & \makecell{5.32\\5.52} & \makecell{6.41\\6.64} & \makecell{7.80\\8.04} 				\comment{Our} & \makecell{\bf3.98\\ \bf4.00} & \makecell{\bf5.02\\ \bf5.07} & \makecell{\bf6.38\\ \bf6.47} & \makecell{\bf7.99\\ \bf8.12} & \makecell{\bf9.73\\ \bf9.80} 					\comment{Fading_1.0} \comment{BP} & \makecell{2.86\\2.87} & \makecell{3.11\\3.13} & \makecell{3.41\\3.43} & \makecell{3.77\\3.81} & \makecell{4.16\\4.24} 				\comment{Our} & \makecell{\bf3.05\\ \bf3.06} & \makecell{\bf3.37\\ \bf3.38} & \makecell{\bf3.73\\ \bf3.75} & \makecell{\bf4.12\\ \bf4.15} & \makecell{\bf4.53\\ \bf4.56} 						\comment{Bursting} \comment{BP} & \makecell{3.26\\3.27} & \makecell{3.85\\3.91} & \makecell{4.58\\4.72} & \makecell{5.44\\5.67} & \makecell{6.43\\6.78} 					\comment{Our} & \makecell{\bf3.42\\ \bf3.42} & \makecell{\bf4.17\\ \bf4.21} & \makecell{\bf5.18\\ \bf5.27} & \makecell{\bf6.40\\ \bf6.56} & \makecell{\bf7.75\\ \bf8.01} \\ \midrule

PGE2(64,32)     \comment{AWGN} \comment{BP} & \makecell{3.78\\3.78} & \makecell{4.38\\4.38} & \makecell{5.12\\5.13} & \makecell{6.04\\6.04} & \makecell{7.17\\7.16} \comment{Our} & \makecell{\bf3.84\\ \bf3.84} & \makecell{\bf4.45\\ \bf4.44} & \makecell{\bf5.19\\ \bf5.19} & \makecell{\bf6.10\\ \bf6.10} & \makecell{\bf7.22\\ \bf7.23} \comment{Fading_1.0} \comment{BP} & \makecell{3.74\\3.74} & \makecell{4.08\\4.08} & \makecell{4.44\\4.44} & \makecell{4.81\\4.81} & \makecell{5.20\\5.20} \comment{Our} & \makecell{\bf3.75\\ \bf3.75} & \makecell{\bf4.10\\ \bf4.10} & \makecell{\bf4.46\\ \bf4.47} & \makecell{\bf4.85\\ \bf4.85} & \makecell{\bf5.24\\ \bf5.24} \comment{Bursting} \comment{BP} & \makecell{\bf3.54\\ \bf3.54} & \makecell{4.07\\4.06} & \makecell{4.69\\4.69} & \makecell{5.43\\5.43} & \makecell{6.29\\6.29} \comment{Our} & \makecell{\bf3.54\\ \bf3.54} & \makecell{\bf4.07\\ \bf4.06} & \makecell{\bf4.70\\ \bf4.69} & \makecell{\bf5.43\\ \bf5.44} & \makecell{\bf6.30\\ \bf6.30} \\ \midrule

PGE5(64,32)     \comment{AWGN} \comment{BP} & \makecell{4.41\\4.78} & \makecell{6.02\\6.63} & \makecell{8.20\\9.06} & \makecell{10.95\\ \bf12.30} & \makecell{14.40\\ \bf15.75} \comment{Our} & \makecell{\bf4.82\\ \bf5.26} & \makecell{\bf6.53\\ \bf7.13} & \makecell{\bf8.73\\ \bf9.48} & \makecell{\bf11.56\\12.20} & \makecell{\bf14.41\\14.90} \comment{Fading_1.0} \comment{BP} & \makecell{4.56\\4.98} & \makecell{5.63\\6.19} & \makecell{6.86\\7.52} & \makecell{8.31\\9.02} & \makecell{9.78\\10.76} \comment{Our} & \makecell{\bf5.09\\ \bf5.67} & \makecell{\bf6.22\\ \bf6.96} & \makecell{\bf7.48\\ \bf8.34} & \makecell{\bf8.82\\ \bf9.85} & \makecell{\bf10.16\\ \bf11.24} \comment{Bursting} \comment{BP} & \makecell{3.84\\4.13} & \makecell{5.18\\5.68} & \makecell{6.97\\7.75} & \makecell{9.34\\ \bf10.19} & \makecell{\bf12.25\\ \bf13.40} \comment{Our} & \makecell{\bf4.19\\ \bf4.55} & \makecell{\bf5.59\\ \bf6.12} & \makecell{\bf7.41\\ \bf8.06} & \makecell{\bf9.51\\10.13} & \makecell{11.84\\12.27} \\ \midrule

  PGE10(64,32)    \comment{AWGN} \comment{BP} & \makecell{3.08\\3.20} & \makecell{3.98\\4.27} & \makecell{5.17\\5.77} & 
  \makecell{6.70\\7.67} & \makecell{8.49\\9.72} \comment{Our} & \makecell{\bf4.21\\ \bf4.62} & \makecell{\bf5.56\\ \bf6.25} & \makecell{\bf7.22\\ \bf8.28} & \makecell{\bf9.13\\ \bf10.59} & \makecell{\bf11.11\\ \bf12.84} \comment{Fading_1.0} \comment{BP} & \makecell{2.96\\3.08} & \makecell{3.52\\3.71} & \makecell{4.18\\4.47} & \makecell{4.95\\5.30} & \makecell{5.81\\6.24} \comment{Our} & \makecell{\bf4.16\\ \bf4.60} & \makecell{\bf5.02\\ \bf5.60} & \makecell{\bf6.00\\ \bf6.72} & \makecell{\bf7.11\\ \bf7.90} & \makecell{\bf8.33\\ \bf9.24} \comment{Bursting} \comment{BP} & \makecell{2.75\\2.82} & \makecell{3.48\\3.67} & \makecell{4.47\\4.90} & \makecell{5.75\\6.46} & \makecell{7.28\\8.26} \comment{Our} & \makecell{\bf3.30\\ \bf3.57} & \makecell{\bf4.26\\ \bf4.73} & \makecell{\bf5.50\\ \bf6.22} & \makecell{\bf7.01\\ \bf8.01} & \makecell{\bf8.80\\ \bf10.06} \\

\bottomrule
	\end{tabular}
	}
\end{table}

\section{Line Search Optimization}
\label{app:lso}
\begin{figure}[h]
% \vspace{1em}
\centering
\noindent  \begin{tabular}{@{}cccc@{}}
        \includegraphics[trim={10 0 10 0},clip, width=0.245\linewidth]{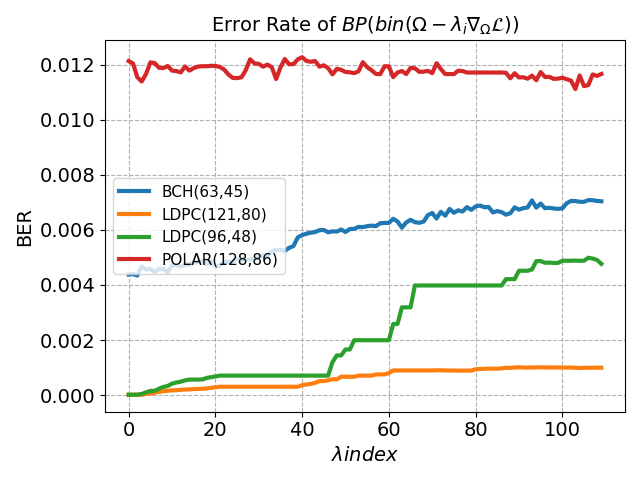}&
        \includegraphics[trim={10 0 10 0},clip, width=0.245\linewidth]{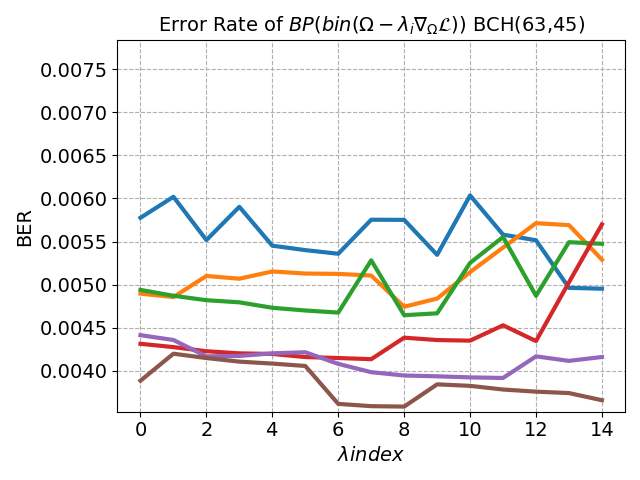}&
        \includegraphics[trim={10 0 10 0},clip, width=0.245\linewidth]{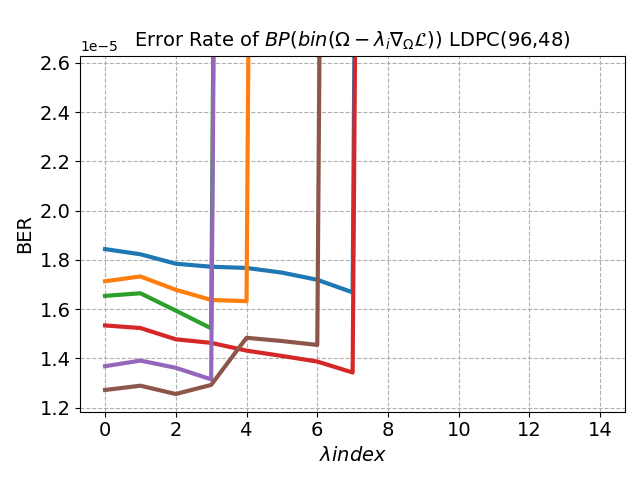}&
        \includegraphics[trim={10 0 10 0},clip, width=0.245\linewidth]{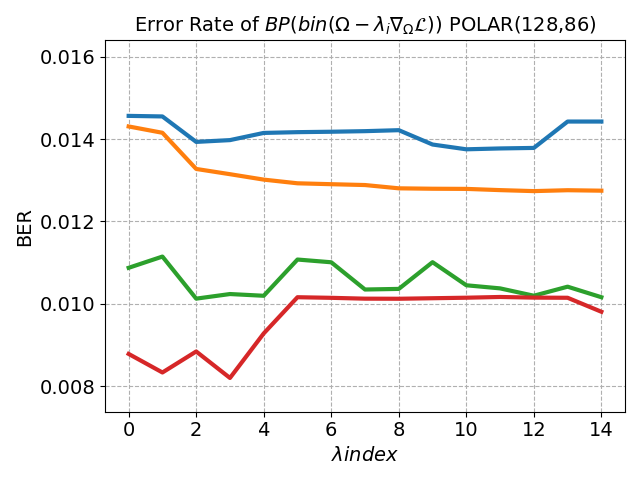}\\
        (a) & (b) & (c) & (d) 
      \end{tabular}
%   \caption{Comparison of the self-attention mechanisms}
  \caption{BER in function of the step size index $i$ on AWGN channel. (a) Averaged BER over the optimization iterations for 4 codes. (b,c,d) BER per optimization iteration for the first 5 optimization iterations and the first 10 indices for three different codes. 
  Here $\lambda_{0}=0$ denotes the original BER.}
%  {\color{black}YOU SEE THE FUNCTION $BER(\lambda_{i})$ for different $i$s. The values of lambda are irrelevant (at least for showing the non-convexity and proximity to the origin at each iteration) between codes, only the index.}
%  }
\label{fig:ls_range}
\end{figure}

In Figure \ref{fig:ls_range} we provide visualizations of the line search procedure.
We provide BER with respect to the step size $\lambda_{i}$ indexed by $i$ ($\lambda_{0}\equiv 0$).
We can observe the high non-convexity of the objective, with the presence of several local minima. We can also notice the proximity of the optimum to the current parity-check estimate (i.e., $\lambda_{0}$).
%%%%%%%%%%%%%%%%%%%%%%%%%%%%%%%%%%%%%%%%%%%%%%%%%%%%%%%%%%%%%%%%%%%%%%%%%%%%%%%%%%%%%%%%%%%%%%%%%%%%%%%%%%%%%%%%%%%%%
%%%%%%%%%%%%%%%%%%%%%%%%%%%%%%%%%%%%%%%%%%%%%%%%%%%%%%%%%%%%%%%%%%%%%%%%%%%%%%%%%%%%%%%%%%%%%%%%%%%%%%%%%%%%%%%%%%%%%
%%%%%%%%%%%%%%%%%%%%%%%%%%%%%%%%%%%%%%%%%%%%%%%%%%%%%%%%%%%%%%%%%%%%%%%%%%%%%%%%%%%%%%%%%%%%%%%%%%%%%%%%%%%%%%%%%%%%%
%%%%%%%%%%%%%%%%%%%%%%%%%%%%%%%%%%%%%%%%%%%%%%%%%%%%%%%%%%%%%%%%%%%%%%%%%%%%%%%%%%%%%%%%%%%%%%%%%%%%%%%%%%%%%%%%%%%%%
%%%%%%%%%%%%%%%%%%%%%%%%%%%%%%%%%%%%%%%%%%%%%%%%%%%%%%%%%%%%%%%%%%%%%%%%%%%%%%%%%%%%%%%%%%%%%%%%%%%%%%%%%%%%%%%%%%%%%
%%%%%%%%%%%%%%%%%%%%%%%%%%%%%%%%%%%%%%%%%%%%%%%%%%%%%%%%%%%%%%%%%%%%%%%%%%%%%%%%%%%%%%%%%%%%%%%%%%%%%%%%%%%%%%%%%%%%%
%%%%%%%%%%%%%%%%%%%%%%%%%%%%%%%%%%%%%%%%%%%%%%%%%%%%%%%%%%%%%%%%%%%%%%%%%%%%%%%%%%%%%%%%%%%%%%%%%%%%%%%%%%%%%%%%%%%%%
%%%%%%%%%%%%%%%%%%%%%%%%%%%%%%%%%%%%%%%%%%%%%%%%%%%%%%%%%%%%%%%%%%%%%%%%%%%%%%%%%%%%%%%%%%%%%%%%%%%%%%%%%%%%%%%%%%%%%
%%%%%%%%%%%%%%%%%%%%%%%%%%%%%%%%%%%%%%%%%%%%%%%%%%%%%%%%%%%%%%%%%%%%%%%%%%%%%%%%%%%%%%%%%%%%%%%%%%%%%%%%%%%%%%%%%%%%%

\section{Convergence Rate}
\label{app:convergence}
\begin{figure}[h]
% \vspace{1em}
\centering
\noindent  \begin{tabular}{@{}cccc@{}}
        \includegraphics[trim={10 0 10 0},clip, width=0.245\linewidth]{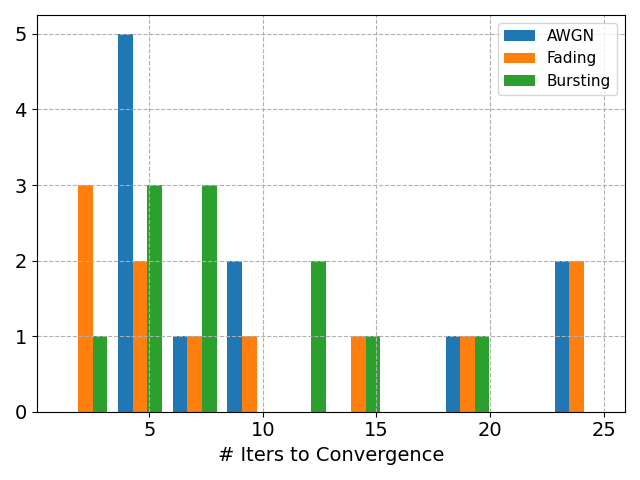}&
        \includegraphics[trim={10 0 10 0},clip, width=0.245\linewidth]{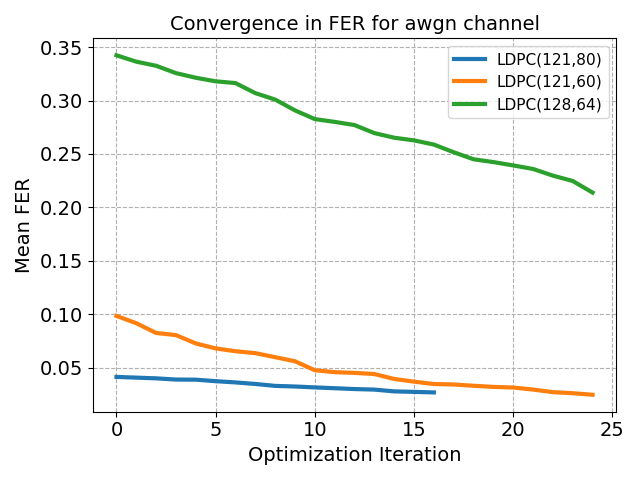}&
        \includegraphics[trim={10 0 10 0},clip, width=0.245\linewidth]{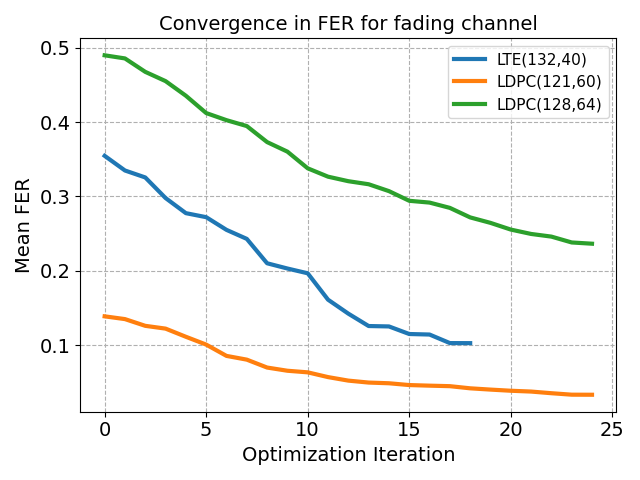}&
        \includegraphics[trim={10 0 10 0},clip, width=0.245\linewidth]{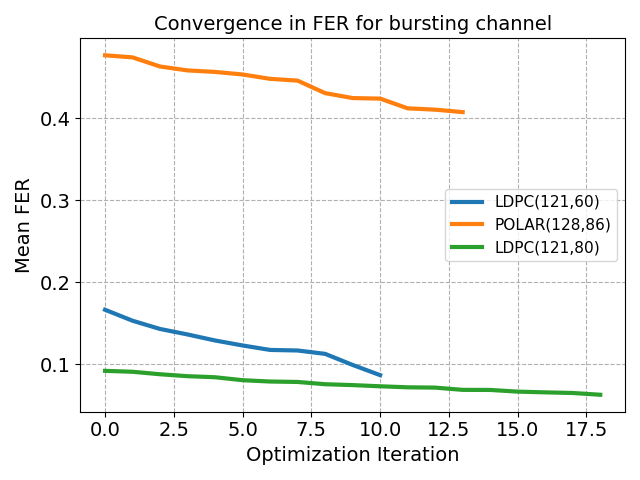}\\
        (a) & (b) & (c) & (d) 
      \end{tabular}
%   \caption{Comparison of the self-attention mechanisms}
  \caption{(a) Histogram of the number of required iterations until convergence. (b) Convergence rate of the Frame Error Rate for three codes on (b) AWGN, (c) fading, and (d) bursting channel. We selected the three codes with the largest number of iterations. The FER is averaged over all the tested $E_{b}/N_{0} = \{3, \dots 7\}$ range.}
\label{fig:convergence}
\end{figure}

In Figure \ref{fig:convergence} we provide statistics on the number of optimization iterations for convergence (a). We also provide (b,c,d) typical convergence. We can observe that the framework typically converges within a few iterations and that the loss decreases monotonically. %Opting for non-descent optimization is also an alternative but is left for future work since the stopping criterion is ill-defined and the number of optimization steps is larger.

%%%%%%%%%%%%%%%%%%%%%%%%%%%%%%%%%%%%%%%%%%%%%%%%%%%%%%%%%%%%

\section{Impact on other BP Variants}
\label{app:bp_variant}
%{\color{red}IN THE TABLE IT IS NOT BP AND OURS BUT ORIGINAL CODE AND OURS THE OTHER METHODS REQUIRE REFERENCES IN THE TEXT. THE DESCRIPTION OF THE RESULTS NEEDS TO BE MORE DETAILED. WHAT IS - IN THE TABLES? RESULTS THAT ARE STILL RUNNING?}
%{\color{black}MODIFIED. PLEASE SEE black IN TABLE. I WILL BOLD BEST LATER FOR MORE CLARITY.}
In Table \ref{tab:min_sum} we provide the performance of the learned code on the more efficient Min-Sum approximation of the Sum-Product algorithm. We can observe that the codes learned with BP consistently outperform the performance of the Min-Sum approximation as well. For some codes, the training range may need to be adjusted.
{We note our method can be applied to neural BP decoders as well.}
The direct optimization over BP approximations and augmentations is left for future work.  

\begin{table}[t]
    \centering
    \caption{
    A comparison of the negative natural logarithm of Bit Error Rate (BER) for five normalized SNR values of our method applied on the Min-Sum BP algorithm. {\color{black} $NE=$ no errors spotted under the testing limits.}}
    \label{tab:min_sum}
    % \setlength\tabcolsep{1.5pt}
%    \smallskip
% \vspace{-2mm}
    \resizebox{0.99985\textwidth}{!}{%
    \begin{tabular}{@{}lc@{~}c@{~}c@{~}c@{~}cc@{~}c@{~}c@{~}c@{~}cc@{~}c@{~}c@{~}c@{~}cc@{~}c@{~}c@{~}c@{~}c@{}}
    \toprule
        BP Method & \multicolumn{10}{c}{Sum-Product} & \multicolumn{10}{c}{Min-Sum} \\
  %       \cmidrule(lr){2-4}
  %       \cmidrule(lr){5-13}
		% \cmidrule(lr){14-16}
        Method & \multicolumn{5}{c}{Baseline} & \multicolumn{5}{c}{Our} & \multicolumn{5}{c}{Baseline} & \multicolumn{5}{c}{Our}  \\
  %       \cmidrule(lr){2-4}
  %       \cmidrule(lr){5-7}
  %       \cmidrule(lr){8-10}
  %       \cmidrule(lr){11-13}
		% \cmidrule(lr){14-16}
         $E_{b}/N_{0}$ & 3 & 4 & 5 & 6 & 7 & 3 & 4 & 5 & 6 & 7 & 3 & 4 & 5 & 6 & 7 & 3 & 4 & 5 & 6 & 7   \\ 
 \midrule                                                                                                    BCH(63,45)      \comment{Sum-Product} \comment{BP} & \makecell{3.35\\ 3.40} & \makecell{4.06\\ 4.22} & \makecell{4.92\\ 5.24} & \makecell{5.98\\ 6.60} & \makecell{7.39\\ 8.33} \comment{Our} & \makecell{3.48\\ 3.57} & \makecell{4.30\\ 4.49} & \makecell{5.29\\ 5.69} & \makecell{6.51\\ 7.17} & \makecell{8.12\\ 9.17} \comment{Min-Sum} \comment{BP} & \makecell{3.04\\ 3.22} & \makecell{3.79\\ 4.09} & \makecell{4.89\\ 5.41} & \makecell{6.33\\ 7.06} & \makecell{8.13\\ 9.14} \comment{Our} & \makecell{3.21\\ 3.40} & \makecell{4.09\\ 4.44} & \makecell{5.32\\ 5.89} & \makecell{6.84\\ 7.60} & \makecell{8.65\\ 10.00} \\ \midrule

CCSDS(128,64)   \comment{Sum-Product} \comment{BP} & \makecell{4.32\\ 4.82} & \makecell{6.47\\ 7.30} & \makecell{9.62\\ 10.70} & \makecell{13.80\\ 15.50} & \makecell{18.40\\ 17.90} \comment{Our} & \makecell{4.44\\ 4.99} & \makecell{6.66\\ 7.57} & \makecell{9.73\\ 11.00} & \makecell{13.60\\ 15.60} & \makecell{18.30\\ $NE$} \comment{Min-Sum} \comment{BP} & \makecell{4.21\\ 4.76} & \makecell{6.62\\ 7.66} & \makecell{10.40\\ 12.20} & \makecell{15.10\\ 17.70} & \makecell{19.40\\ $NE$} \comment{Our} & \makecell{4.35\\ 4.97} & \makecell{6.82\\ 8.03} & \makecell{10.50\\ 12.30} & \makecell{15.00\\ 17.40} & \makecell{21.00\\ $NE$} \\ \midrule

LDPC(32,16)     \comment{Sum-Product} \comment{BP} & \makecell{3.46\\ 3.61} & \makecell{4.39\\ 4.66} & \makecell{5.60\\ 6.07} & \makecell{7.20\\ 7.87} & \makecell{9.23\\ 10.30} \comment{Our} & \makecell{3.62\\ 3.80} & \makecell{4.59\\ 4.91} & \makecell{5.83\\ 6.36} & \makecell{7.45\\ 8.16} & \makecell{9.52\\ 10.70} \comment{Min-Sum} \comment{BP} & \makecell{3.36\\ 3.55} & \makecell{4.38\\ 4.69} & \makecell{5.75\\ 6.21} & \makecell{7.65\\ 8.21} & \makecell{10.10\\ 10.90} \comment{Our} & \makecell{3.53\\ 3.74} & \makecell{4.61\\ 4.93} & \makecell{6.01\\ 6.50} & \makecell{7.92\\ 8.44} & \makecell{10.10\\ 11.10} \\ \midrule

LDPC(96,48)     \comment{Sum-Product} \comment{BP} & \makecell{4.70\\ 5.20} & \makecell{6.73\\ 7.55} & \makecell{9.52\\ 10.70} & \makecell{13.20\\ 14.40} & \makecell{17.30\\ 18.50} \comment{Our} & \makecell{5.01\\ 5.70} & \makecell{7.11\\ 8.13} & \makecell{9.92\\ 11.30} & \makecell{13.50\\ 14.70} & \makecell{16.90\\ 17.40} \comment{Min-Sum} \comment{BP} & \makecell{4.71\\ 5.23} & \makecell{6.96\\ 7.89} & \makecell{9.95\\ 11.50} & \makecell{14.20\\ 15.10} & \makecell{18.30\\ 19.10} \comment{Our} & \makecell{4.98\\ 5.68} & \makecell{7.16\\ 8.32} & \makecell{10.10\\ 12.00} & \makecell{14.00\\ 14.80} & \makecell{15.80\\ 15.80} \\ 

		\bottomrule
	\end{tabular}
	}
\end{table}

\section{Improvement Statistics on all the Codes}
\label{app:stats_all_codes}
\begin{figure}[h]
% \vspace{1em}
\centering
\noindent  \begin{tabular}{@{}ccc@{}}
        \includegraphics[trim={0 0 0 0},clip, width=0.33\linewidth]{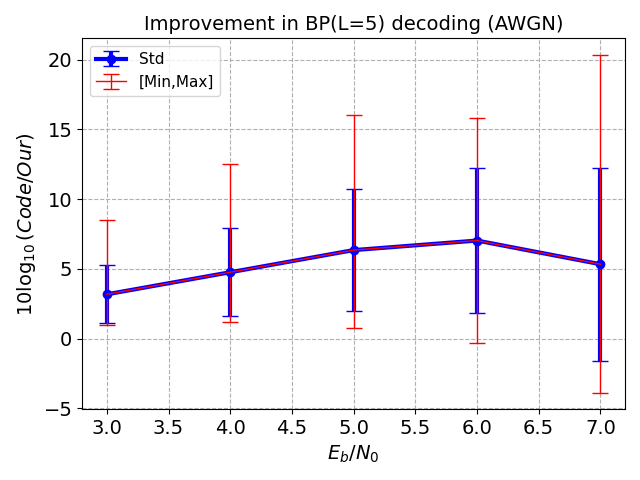}&
        \includegraphics[trim={0 0 0 0},clip, width=0.33\linewidth]{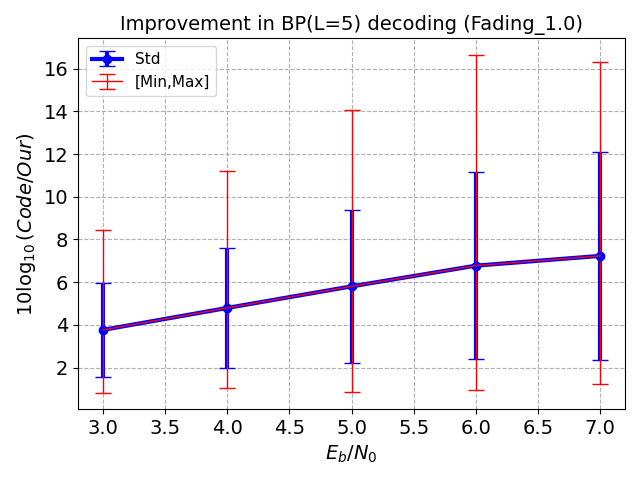}&
        \includegraphics[trim={0 0 0 0},clip, width=0.33\linewidth]{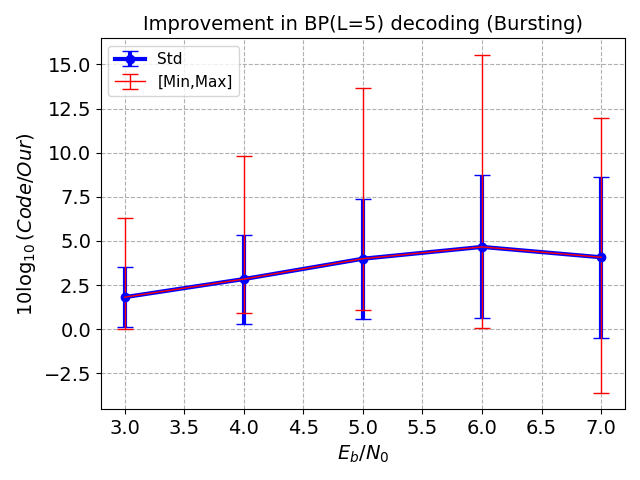}\\
        \includegraphics[trim={0 0 0 0},clip, width=0.33\linewidth]{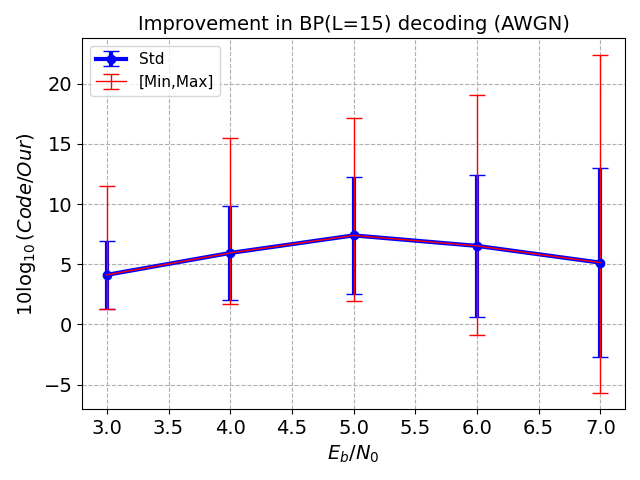}&
        \includegraphics[trim={0 0 0 0},clip, width=0.33\linewidth]{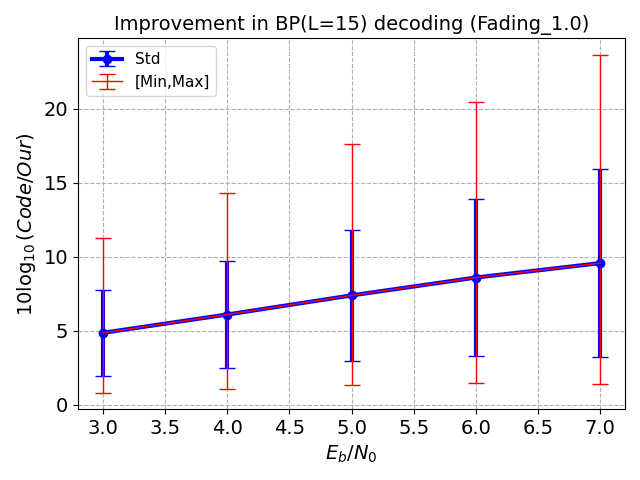}&
        \includegraphics[trim={0 0 0 0},clip, width=0.33\linewidth]{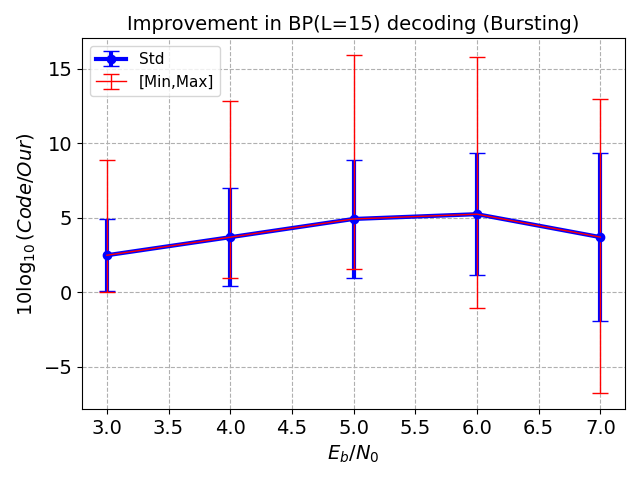}\\
      (a) & (b) & (c)\\
      \end{tabular}
%   \caption{Comparison of the self-attention mechanisms}
  \caption{Statistics of improvement in dB for the  (a) AWGN, (b) fading, and (c) bursting channel on all the codes from Table \ref{tab:main_ber}. 
  We provide the mean and standard deviation as well as the minimum and maximum improvements.}
\label{fig:stats_all}
\end{figure}

We provide in Figure \ref{fig:stats_all} the statistics of improvement on all the codes presented in Table \ref{tab:main_ber}.

\section{More Random Codes}
\label{app:more_rand_codes}
\begin{figure}[h]
% \vspace{1em}
\centering
\noindent  \begin{tabular}{@{}cccc@{}}
        \includegraphics[trim={10 0 10 0},clip, width=0.245\linewidth]{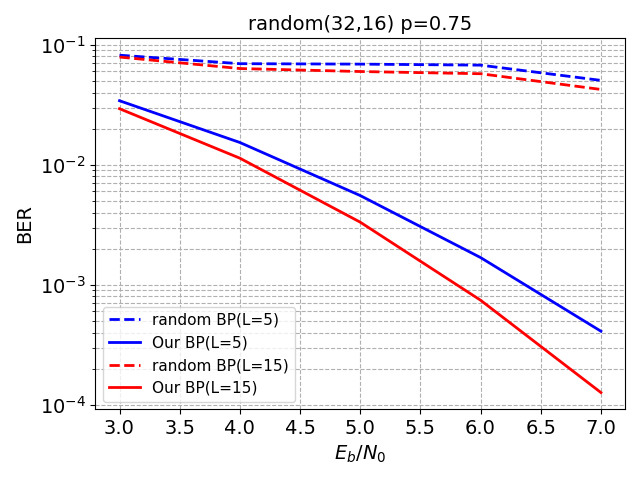}&
        \includegraphics[trim={10 0 10 0},clip, width=0.245\linewidth]{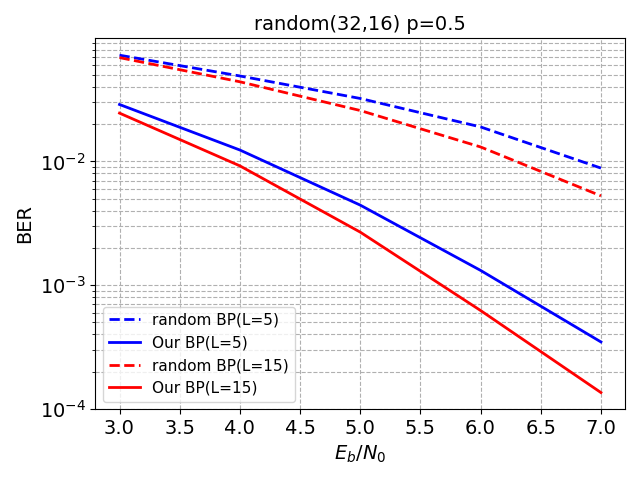}&
        \includegraphics[trim={10 0 10 0},clip, width=0.245\linewidth]{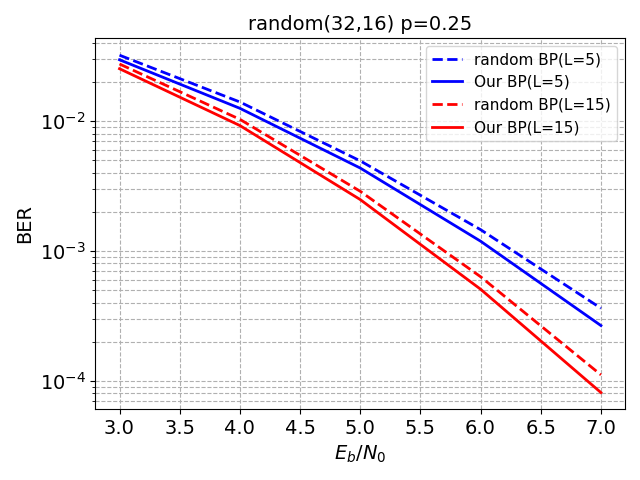}&
        \includegraphics[trim={10 0 10 0},clip, width=0.245\linewidth]{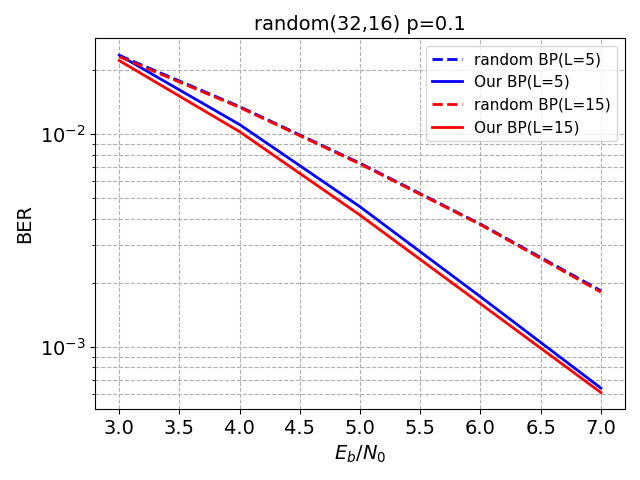} \\
        %%%%%%%%%%%%%%%%%%%%%%%%%%%
        %  \includegraphics[trim={10 0 10 0},clip, width=0.245\linewidth]{Ablations/more_rand_codes/std_form_-1_BER_curves_folder_BP_Yoni_Results_ABLATION6/AWGN/random(64,32)__p_0_75__std_form_False.png}&
        % \includegraphics[trim={10 0 10 0},clip, width=0.245\linewidth]{Ablations/more_rand_codes/std_form_-1_BER_curves_folder_BP_Yoni_Results_ABLATION6/AWGN/random(64,32)__p_0_5__std_form_False.png}&
        % \includegraphics[trim={10 0 10 0},clip, width=0.245\linewidth]{Ablations/more_rand_codes/std_form_-1_BER_curves_folder_BP_Yoni_Results_ABLATION6/AWGN/random(64,32)__p_0_25__std_form_False.png}&
        % \includegraphics[trim={10 0 10 0},clip, width=0.245\linewidth]{Ablations/more_rand_codes/std_form_-1_BER_curves_folder_BP_Yoni_Results_ABLATION6/AWGN/random(64,32)__p_0_1__std_form_False.png} \\
        % %%%%%%%%%%%%%%%%%%%%%%%%%%%
         \includegraphics[trim={10 0 10 0},clip, width=0.245\linewidth]{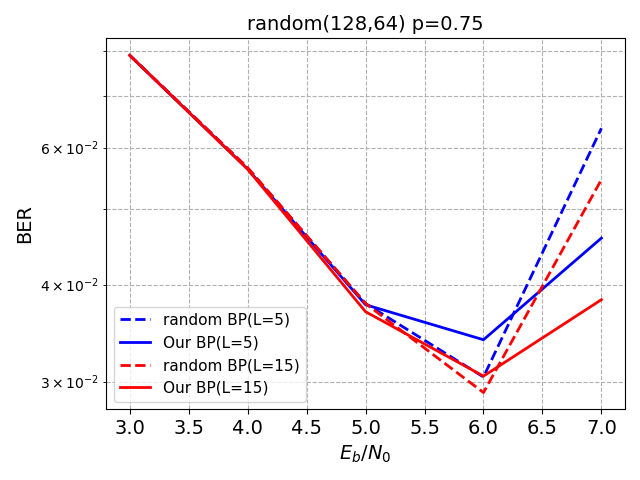}&
        \includegraphics[trim={10 0 10 0},clip, width=0.245\linewidth]{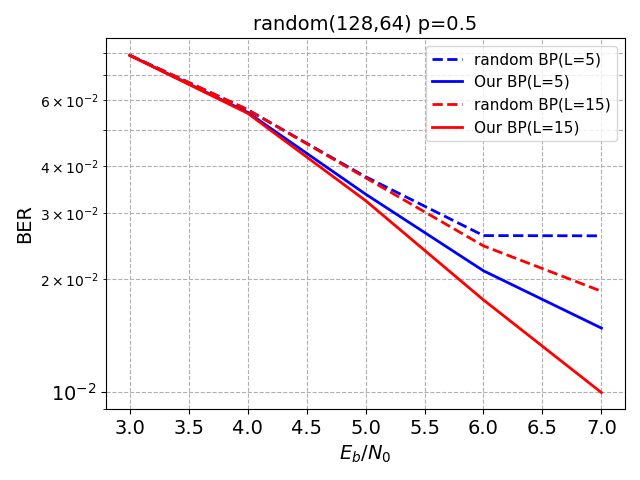}&
        \includegraphics[trim={10 0 10 0},clip, width=0.245\linewidth]{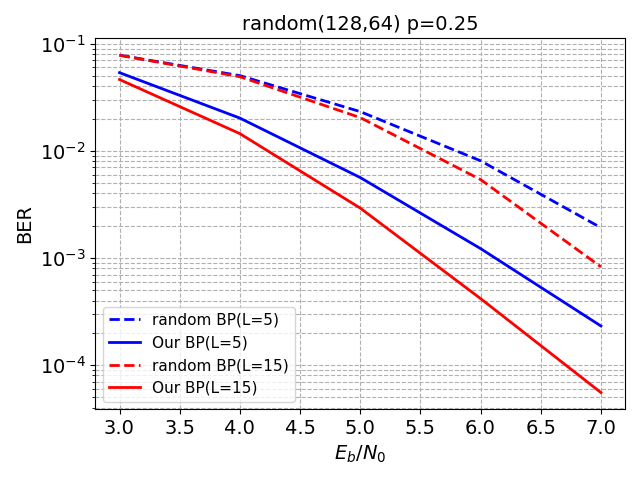}&
        \includegraphics[trim={10 0 10 0},clip, width=0.245\linewidth]{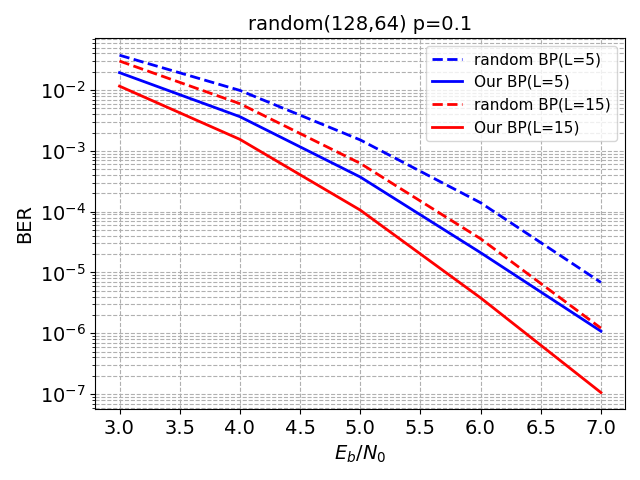} 
      \end{tabular}
%   \caption{Comparison of the self-attention mechanisms}
  \caption{Performance of the method on random codes under different sparsity rate initialization $p$.}
\label{fig:random_codes_reg2}
\end{figure}

\begin{figure}[h]
% \vspace{1em}
\centering
\noindent  \begin{tabular}{@{}cccc@{}}
        \includegraphics[trim={10 0 10 0},clip, width=0.245\linewidth]{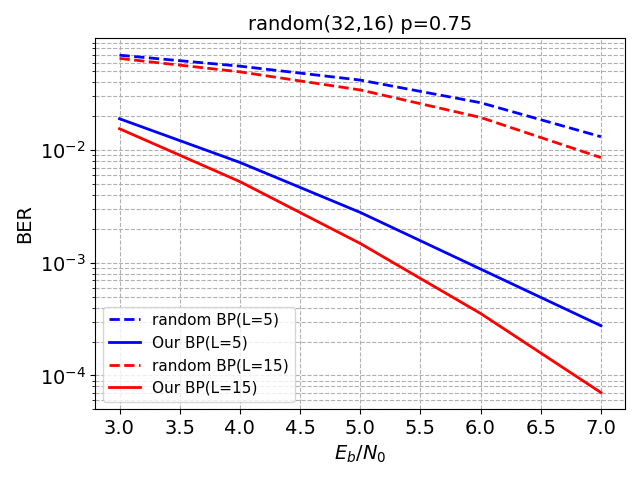}&
        \includegraphics[trim={10 0 10 0},clip, width=0.245\linewidth]{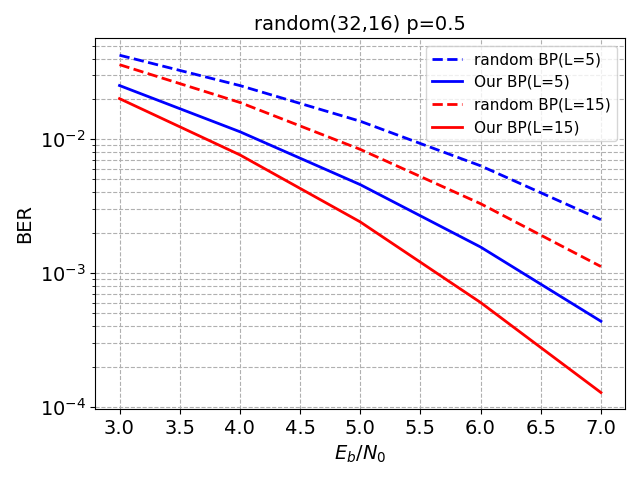}&
        \includegraphics[trim={10 0 10 0},clip, width=0.245\linewidth]{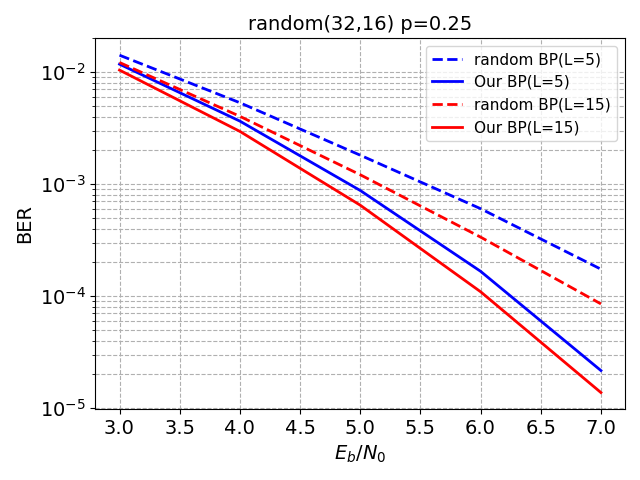}&
        \includegraphics[trim={10 0 10 0},clip, width=0.245\linewidth]{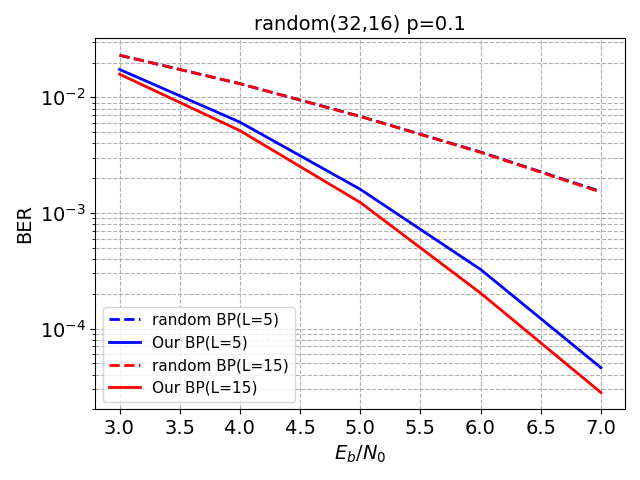} \\
        % %%%%%%%%%%%%%%%%%%%%%%%%%%%
        %  \includegraphics[trim={10 0 10 0},clip, width=0.245\linewidth]{Ablations/more_rand_codes/std_form_1_BER_curves_folder_BP_Yoni_Results_ABLATION6/AWGN/random_64,32)__p_0_75__std_form_True.png}&
        % \includegraphics[trim={10 0 10 0},clip, width=0.245\linewidth]{Ablations/more_rand_codes/std_form_1_BER_curves_folder_BP_Yoni_Results_ABLATION6/AWGN/random_64,32)__p_0_5__std_form_True.png}&
        % \includegraphics[trim={10 0 10 0},clip, width=0.245\linewidth]{Ablations/more_rand_codes/std_form_1_BER_curves_folder_BP_Yoni_Results_ABLATION6/AWGN/random_64,32)__p_0_25__std_form_True.png}&
        % \includegraphics[trim={10 0 10 0},clip, width=0.245\linewidth]{Ablations/more_rand_codes/std_form_1_BER_curves_folder_BP_Yoni_Results_ABLATION6/AWGN/random_64,32)__p_0_1__std_form_True.png} \\
        %%%%%%%%%%%%%%%%%%%%%%%%%%%
         \includegraphics[trim={10 0 10 0},clip, width=0.245\linewidth]{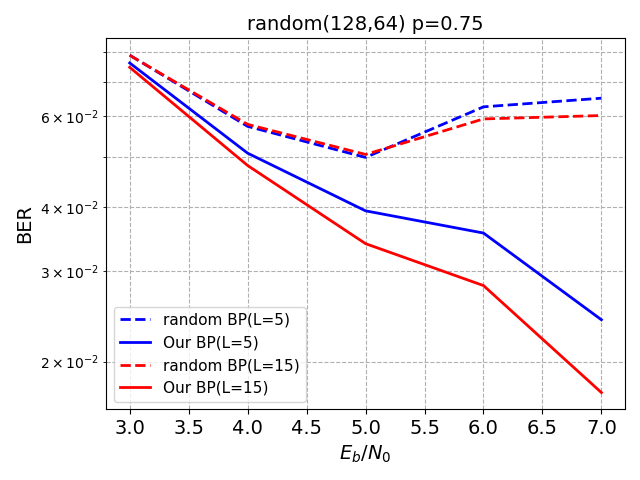}&
        \includegraphics[trim={10 0 10 0},clip, width=0.245\linewidth]{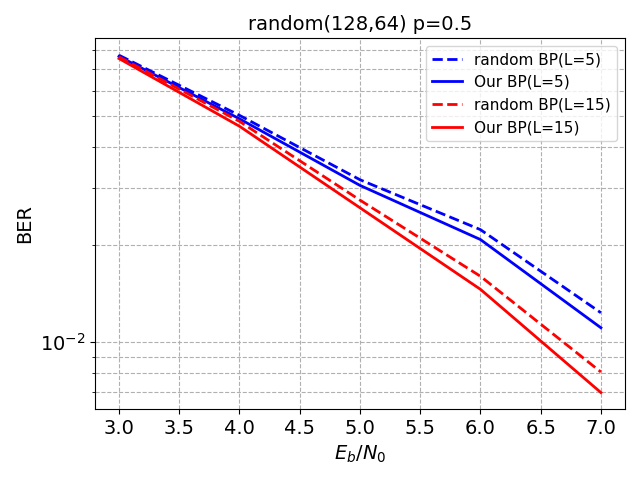}&
        \includegraphics[trim={10 0 10 0},clip, width=0.245\linewidth]{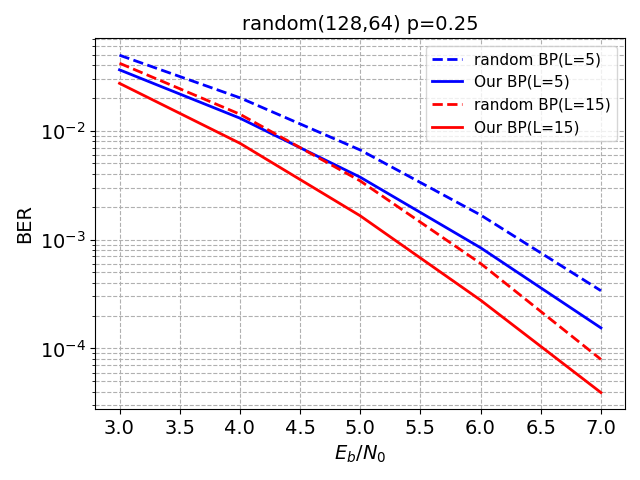}&
        \includegraphics[trim={10 0 10 0},clip, width=0.245\linewidth]{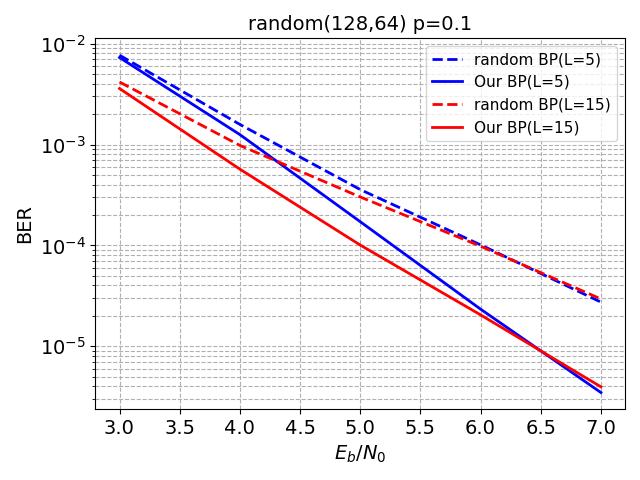} 
      \end{tabular}
%   \caption{Comparison of the self-attention mechanisms}
  \caption{Performance of the method on constrained systematic random codes under different sparsity rate initialization $p$ on the AWGN channel.}
\label{fig:random_codes_std2}
\end{figure}

{\color{black}
We provide in Figure \ref{fig:random_codes_reg2} the performance of the proposed method on random codes initialized with different sparsity rates on different lengths.
We also provide in Figure \ref{fig:random_codes_std2} the performance of the proposed method on constrained systematic random codes initialized with different sparsity rates on different lengths.
}

\section{Comparison with Genetic Algorithm}
\label{app:other_baselines}
% \subsection{Comparison with Genetic Algorithm}
We provide in table \ref{tab:main_r1} a comparison with the genetic algorithm of \cite{elkelesh2019decoder}.
We note that the method requires 230 offspring/code evaluations per iteration, with 300 iterations (Fig. 7 in \citep{elkelesh2019decoder}) or even an infinite loop (cf. the provided MATLAB code). 
Our algorithm is tested on 50 line-search steps as described in in the paper on 2 to 25 iterations (cf. App. \ref{app:convergence}), which means that \cite{elkelesh2019decoder} requires approximately 25 to 313 times more computations than our proposed method. 
The performance presented are for 75 and 150 iterations of the algorithm, representing around 70 and 140 times slower performance than our approach, respectively, while they remain below our performance.
We note here, as described in the paper, that combining the methods by allowing the perturbation of the parity-check matrix at a local minima may allow the discovery of other better local optimum.

\begin{table}[h]
    \centering
    \caption{
    A comparison of the negative natural logarithm of Bit Error Rate (BER) for several normalized SNR values of our method with classical codes. Higher is better. BP results are provided for 5 iterations in the first row and 15 in the second row. The best results are in bold. The training range is defined as $E_{b}/N_{0}=\{5\}$. GA denotes the genetic algorithm of \cite{elkelesh2019decoder} with $k$ iterations.}
    \label{tab:main_r1}
    % \setlength\tabcolsep{1.5pt}
%    \smallskip
% \vspace{-2mm}
    % \resizebox{0.89985\textwidth}{!}{%
    \resizebox{0.9\textwidth}{!}{%
    \begin{tabular}{@{}lc@{~}c@{~}cc@{~}c@{~}cc@{~}c@{~}cc@{~}c@{~}c}
    \toprule
        Channel & \multicolumn{12}{c}{AWGN} \\
  %       \cmidrule(lr){2-4}
  %       \cmidrule(lr){5-13}
		% \cmidrule(lr){14-16}
        Method & \multicolumn{3}{c}{BP} & \multicolumn{3}{c}{Our} & \multicolumn{3}{c}{GA $k=75$} & \multicolumn{3}{c}{GA $k=150$} \\
  %       \cmidrule(lr){2-4}
  %       \cmidrule(lr){5-7}
  %       \cmidrule(lr){8-10}
  %       \cmidrule(lr){11-13}
		% \cmidrule(lr){14-16}
         $E_{b}/N_{0}$ & 4 & 5 & 6  & 4 & 5 & 6  & 4 & 5 & 6 & 4 & 5 & 6   \\ 
         \midrule                                                                                                                                                                                                                                                                                                                                                                                                                                      
		%%%%%%%%%%%%%%%%%%%									
CCSDS(128,64)   	\comment{AWGN} \comment{BP} & \makecell{6.46\\7.32} & \makecell{9.61\\10.83} & \makecell{13.99\\15.43} 		 		
\comment{Our} & \makecell{\bf7.34\\ \bf8.61} & \makecell{\bf10.48\\ \bf12.26} & \makecell{\bf14.37\\ 16.00} 	
\comment{R1} & \makecell{6.86\\7.87} & \makecell{9.95\\11.31} & \makecell{13.38\\15.56}	
\comment{R1 2} & \makecell{7.09\\8.23} & \makecell{10.40\\11.79} & \makecell{14.08\\ \bf16.04}\\

\bottomrule
	\end{tabular}
	}
\end{table}

\clearpage
\end{document}